\documentclass[10pt]{article}
\usepackage{graphicx}
\usepackage{amssymb}
\usepackage{multirow}
\usepackage{subfig}
\usepackage[margin=3.5cm]{geometry}

\newcommand{\bd}{\begin{displaymath}}
\newcommand{\ed}{\end{displaymath}}
\newcommand{\be}{\begin{equation}}

\newcommand{\ee}{\end{equation}}
\newcommand{\ben}{\begin{eqnarray}}
\newcommand{\een}{\end{eqnarray}}

\newcommand{\vbc}{v_{\rm bc}}
\newcommand{\vbg}{v^{\rm bg}}
\newcommand{\vc}{v_{\rm c}}
\newcommand{\vb}{v_{\rm b}}

\def\sqr#1#2{{\vcenter{\vbox{\hrule height.#2pt
         \hbox{\vrule width.#2pt height#1pt \kern#1pt
            \vrule width.#2pt}
         \hrule height.#2pt}}}}

\begin{document}

\title{Supersonic Relative Velocity between Dark Matter and Baryons: A Review}
\author{Anastasia Fialkov\footnote{E-mail: anastasia.fialkov@phys.ens.fr}\\ \\
\it{\small Departement de Physique, Ecole Normale Superieure,} \\ \it{\small Centre National de la Recherche Scientifique, } \\ \it{\small 24 rue Lhomond, 75005 Paris, France.}}
\date{}
\maketitle

\begin{abstract}

Our understanding of astrophysical and cosmological phenomena in recent years has improved enormously thanks to precision measurements of various cosmic signals such as Cosmic Microwave Background radiation, emission of galaxies and dust, spectral lines attributed to various elements etc. Despite this, our knowledge at intermediate redshifts ($10<z<1100$) remains fragmentary and incomplete, and as a consequence, various physical processes happening between the epochs of hydrogen  recombination and reionization remain still highly unconstrained. Moreover, some important fragments of the theoretical description that are less decisive for the Universe today, but that had an important impact at intermediate redshifts, have been omitted in some of the studies concerning the Universe at high redshifts. One such neglected phenomenon, only recently brought to the attention of the community  by D. Tseliakhovich and C. M. Hirata (2010) \cite{Tseliakhovich:2010}, is the fact that after hydrogen recombination the large-scale baryons and dark matter fluctuations had supersonic relative velocities: the central topic of this review. The relative velocities between dark matter and baryons formally introduce a second order effect on the standard results and thus have been  neglected in the framework of linear theory. However, when properly considered, the velocities yield a non-perturbative contribution to the growth of structures which is then inherited by the majority of cosmic signals coming from redshifts above $z \sim 10$, and in certain cases may even propagate to various low-redshift observables such as the Baryon Acoustic Oscillations measured from the distribution of  galaxies. At higher redshifts, the supersonic velocities have thus strong impact affecting the abundance of M $\sim 10^6$ M$_\odot$ halos in an inhomogeneous way, hindering the formation of first stars, leaving traces in the redshifted 21-cm signal of neutral hydrogen, as well as having other important contributions at high redshifts all of which we review in this manuscript.

\end{abstract}

\tableofcontents

\section{Overview}

Understanding the early Universe and building a coherent picture of the evolution of cosmic structures that grew from tiny initial perturbations (which, for instance,  may have been  produced during cosmological inflation \cite{Guth:1981, Linde:1982}) to become the stars and galaxies that we observe with our telescopes is one of the major research goals of cosmology  today. At present,  many  unprecedentedly large observatories and telescopes are being projected, developed and realized,   with the goal to test existing theories at previously inaccessible limits, explain seemingly inconsistent observations and, possibly, discover new unexpected phenomena. Among such experiments are  the Square Kilometer Array\footnote{www.skatelescope.org} (SKA), intended for observations in the radio range of the electromagnetic spectrum, one of the main goals of which is to map the redshifted  21-cm signal of neutral hydrogen from the epoch of primordial star formation; the  James Webb Space Telescope\footnote{www.jwst.nasa.gov} (JWST), optimized for the observations in the infrared to study the birth and evolution of galaxies (including the most distant ones) as well as formation of stars and planets;  Euclid\footnote{www.sci.esa.int$/$euclid$/$}, whose main objective is surveying galaxies and their redshifts to constrain dark energy;  the Polarized Radiation Imaging and Spectroscopy Mission\footnote{www.prism-mission.or} (PRISM), which is expected to probe cosmic structures and radiation on the microwave sky,  and many other probes. This observational effort is expected to expand our horizons and to shed light on some of the crucial questions related to the state of the Universe at epochs between hydrogen recombination and the end of the epoch of reionization. Among the open questions that these experiment aim on answering, are the nature of the first yet unobserved stars, the character radiative backgrounds which heated and re-ionizaed the pristine gas, the mechanism through which observed supermassive black holes \cite{Fan:2006, Mortlock:2011,Venemans:2013} were formed, and other important subjects. 

Extracting the knowledge from the future science-rich data sets is expected to be a difficult task, partially due to the lack of theoretical modeling of the high-redshift Universe. In fact, available theoretical predictions are not yet at the required level of precision to relate observed signals to a particular set of astrophysical and cosmological parameters, though rapid progress is being made in this field. Moreover, some of the crucial components of the theoretical description have sometimes been  simply omitted form consideration, as is the case of the supersonic  relative motion between dark matter and gas, which was only recently acknowledged for by  D. Tseliakhovich and C. M. Hirata (2010) \cite{Tseliakhovich:2010}. This non-linear effect, which was incorrectly neglected as second order  until recently, appears to have a non-negligible impact on formation of first cosmic structures including dark matter halos, galaxies, stars and black holes, as well as on their products (e. g., radiative backgrounds emitted by stellar populations). As a result, the  supersonic relative motion is prone to affect theoretical expectations relevant for a large part of future observations. 

Seeded at the time of hydrogen  recombination at redshift $z\sim 1100$ \cite{Ade:2013}, the relative velocities decay with redshift and are expected to be negligible by the epoch of reionization at $z\sim 11$ \cite{Ade:2013}. Therefore, the strongest impact of the velocities is expected to be on the formation of early structures such as the first stars \cite{Maio:2011, Stacy:2011, Greif:2011, OLeary:2012, Fialkov:2012, Naoz:2012, Naoz:2013}. However, residual  effects may also be found in late time phenomena, such as present day galaxies \cite{Yoo:2011, Bovy:2013, Yoo:2013, Slepian:2014} and globular clusters \cite{Naoz:2014},  and the B-mode polarization \cite{Ferraro:2012} of the Cosmic Microwave Background (CMB), thus possibly altering predictions for all the experiments listed above.  
 
The large scale cosmic flows of gas relatively to the dark matter potential wells originate from different evolution scenarios for the perturbations in dark matter and baryons at early times: before the Universe expanded enough to allow first atoms to recombine, photons,  electrons and protons  were tightly coupled together  and formed a hot dense plasma in which acoustic waves propagated. Unlike baryons that could not cluster due to the coupling to photons, dark matter particles  were  undergoing gravitational collapse ever since the matter-radiation equality at $z\sim 3400$ \cite{Ade:2013}, thus developing  significant infall velocities by the time of decoupling of the CMB photons. By then therefore,  the relative velocities of baryons with respect to dark matter particles were non-negligible, and in addition they kept trace of the Baryon Acoustic Oscillations (BAO), i.e. the magnitude of the relative velocities varied from place to place with the characteristic length scale of $\sim150$ Mpc. Moreover, the sound speed of baryons  dropped  strongly at decoupling, thus making the relative velocities supersonic in average over the Universe \cite{Tseliakhovich:2010}. 

The fact that right after cosmological recombination baryons and dark matter particles had different velocities is not new and has been extensively discussed in literature. The residual velocities of baryons right after recombination and their effect on fluctuations in the baryon density field were mentioned in literature for the first time in 1970 by R. A. Sunyaev and Ya. B. Zeldovich (1970) \cite{Sunyaev:1970} (see also the paper by W. H. Press and E. T. Vishniac \cite{Press:1980}). Moreover, the implications of the baryonic ``velocity overshoot'' effect on the CMB and on the matter transfer function were considered by W. Hu and N. Sugiyama (1996) \cite{Hu:1996}, who also briefly mentioned that there should be an interplay between gravitational collapse and the ``overshoot effect''. However, the fact that the relative motion between baryons and dark matter was highly supersonic right after recombination, as well as the crucial implications of this motion on structure  formation at high redshifts, was only noticed in 2010 by D. Tseliakhovich and C. M. Hirata \cite{Tseliakhovich:2010} and was not discussed previously in literature.

After the supersonic nature of the relative velocities was discovered \cite{Tseliakhovich:2010}, its impact on astrophysics and cosmology was explored using various computational tools which include analytical methods, hydrodynamical simulations and hybrid computational approaches. On the first place, it was shown that as a result of the supersonic relative motion, some second-order terms in fluid equations which describe evolution of baryons and dark matter after recombination become non-perturbative at small scales. Normally ignored in the linear-order analysis, these terms  appear to be important when the formation of first structures at early times is considered \cite{Tseliakhovich:2010}. In the scenario which includes relative velocities, the first collapsed baryonic objects are forced to form in a moving background of dark matter potential wells \cite{Tseliakhovich:2010,  Tseliakhovich:2011, Dalal:2010}. Relative velocities can then be viewed as an anisotropic pressure term (in addition to the hydrostatic pressure of baryons) which  hinders the process of gas accretion by dark matter halos and redistributes gas density within halos \cite{OLeary:2012}.  Consequently, in order to reach high enough gas density and start forming stars, haloes in the presence of relative velocities have to be heavier \cite{Maio:2011, Stacy:2011, Greif:2011, OLeary:2012} than what they would be if no velocity effect was present \cite{Haiman:1996, Tegmark:1997, Machacek:2001, Abel:2002}. Due to this effect, stars form later with an inhomogeneous delay biased by the local value of the relative velocity \cite{Fialkov:2012, Visbal:2012}. However, one should keep in mind that the complete effect of the supersonic motion on star formation and various physical observables depends on  complex astrophysics at high redshifts, which at the moment is not constrained by observations and is difficult to model. For instance, negative feedback to star formation by Lyman-Werner photons, which shut down formation of the first bright objects in the smallest halos \cite{Haiman:1997},  can largely erase the effect of the relative motion on cosmic signals \cite{Fialkov:2013}.

Here we review the role that the relative supersonic motion between gas and dark matter plays in the early Universe.  The review is built as following: in section \ref{Sec:basics} we characterize  the supersonic relative motion in detail. In section \ref{Sec:structure} the effect of relative velocities on cosmic structure is reviewed, including its impact on halo abundance, gas content of halos and  implications on formation of bounded structures such as the first stars and black holes. Next, in section \ref{Sec:observ} we discuss the effect of the relative motion on cosmological signals, focusing in particular on the redshifted  21-cm signal of  neutral hydrogen, which is the only known way to study high redshift Universe at the dawn of star formation and during the cosmic dark ages. We also comment on the effect relative velocities may have on the CMB B-mode polarization, and on the number counts of faint galaxies at high redshifts. For completeness, we must note that  the concept of supersonic relative velocities was recently applied to clustering of neutrinos after they become non-relativistic \cite{Zhu:2013}. However, this subject is out of the scope of this review. 

Throughout this work  we assume that the Universe is described by a $\Lambda$CDM cosmology with parameter values in accordance to what has been recently measured by the Plank satellite \cite{Ade:2013}.  

\section{Origins of the Supersonic Motion}
\label{Sec:basics}

From the observations of the CMB we know that the early Universe was very smooth and isotropic with small perturbations in its energy density.  Thanks to experimental efforts such as  Cosmic Background Explorer (COBE) satellite \cite{COBE} and its successors, the temperature fluctuations (which probe density fluctuations) were measured to be just $\sim 1/10^5$ of the average value of temperature around the sky, with no detected deviations from a Gaussian distribution \cite{Ade:2013b}. Defining the (dimensionless) fluctuations for both dark matter and baryon densities as $\delta \equiv (\rho-\bar \rho)/\bar \rho$ (where $\rho$ represents the local density and $\bar \rho$ the density averaged over the size of the observable Universe), the observed fluctuations in the CMB constrain them to be 
of magnitude $\delta \sim 10^{-5}$. These tiny initial fluctuations in density grew in time as a result of gravitational attraction giving rise to all the variety of cosmic structures that we observe today. 

To describe the time evolution of the dimensionless density fluctuations the following complete set of nonlinear equations is used. It includes  the continuity equations, written for the energy density fluctuations for baryons ($\delta_b$) and cold dark matter ($\delta_c$), the  pressure-less Navier-Stokes equation for cold dark matter, the Navier-Stokes equation with a pressure term for baryons, and the Poisson equation for the gravitational potential $\Phi$:
\begin{equation}
\begin{array}{l}
\displaystyle\frac{\partial\delta_c}{\partial t}+a^{-1} {\bf v}_c\cdot\nabla \delta_c = -a^{-1}(1+
 \delta_c)\nabla\cdot{\bf v_c},\\\\
\displaystyle\frac{\partial\delta_b}{\partial t}+a^{-1} {\bf v}_b\cdot\nabla \delta_b = -a^{-1}(1+
 \delta_b)\nabla\cdot{\bf v_b},\\\\
\displaystyle \frac{\partial {\bf v}_c}{\partial t}+a^{-1} ({\bf v}_c\cdot\nabla){\bf v}_c = -\frac{\nabla \Phi}{a}-H{\bf v}_c,\\\\
\displaystyle\frac{\partial {\bf v}_b}{\partial t}+a^{-1} ({\bf v}_b\cdot\nabla){\bf v}_b = -\frac{\nabla \Phi}{a}-H{\bf v}_b-a^{-1}c_s^2\nabla\delta_b,\\\\
 \displaystyle a^{-2}\nabla^2\Phi = 4\pi G\bar\rho_m\delta_m,
\end{array}
\label{eq:sys1}
\end{equation} 
where 
${\bf \vb}$ and ${\bf \vc}$ are the velocities of  baryon and  dark matter respectively, $a$ is the scale factor,  $H \equiv \dot a/a$ is the Hubble constant, and 
\begin{equation}
\label{eq:cs2}c_s^2 =\frac{ k_B T_K}{\mu }\left(1-\frac{1}{3}\frac{\partial\log T_K}{\partial\log a}\right),
\end{equation}
 is the baryonic sound speed (which in general is non-uniform \cite{Naoz:2005}), with $k_B$ the Boltzmann constant, $T_K$  the kinetic gas  temperature, and $\mu$ the mean molecular weight. These equations (eq. \ref{eq:sys1})	describe the evolution of perturbations of two coupled fluids (baryons and cold dark matter) in the expanding $\Lambda$CDM Universe.

Prior to recombination perturbations in the density of dark matter and baryons evolve independently, in the sense that, being tightly coupled to photons, baryons do not fall into the potential wells of dark matter, and their evolution is well described by linear theory while they are small. Fluctuations in dark matter grow after matter-radiation equality, and by the redshift of hydrogen recombination the dark matter particles acquire significant velocities due to gravitational acceleration, whereas baryons 
are supported by radiative pressure against gravitational collapse. Moreover, they are highly relativistic due to the strong coupling to the radiation field. As a result of the different evolution scenarios before recombination, baryons and dark matter acquire significant relative velocities,  denoted here by ${\bf\vbc} \equiv {\bf \vb} -{\bf \vc}$, the amplitudes of which are denoted by $\vbc$.  

\begin{figure}[t]
{\center
\includegraphics[width=3.6in]{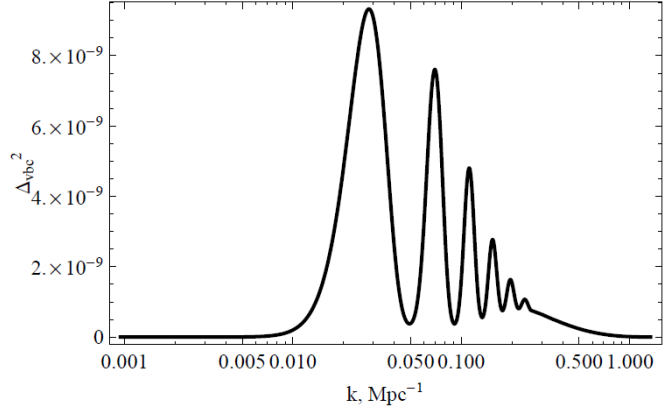}
\caption{\label{fig:vbc} The variance of the perturbations in the relative velocity field per $\log k$, where k is wavenumber in Mpc$^{-1}$. The power spectrum shows clear Baryon Acoustic Oscillations at  $k \sim 0.01-0.5$ Mpc$^{-1}$ scales. At larger wavenumbers ($k>0.5$ Mpc$^{-1}$) the power spectrum drops fast, indicating that the flow is coherent on smaller scales.  The figure is adopted from D. Tseliakhovich and C. M. Hirata (2010) \cite{Tseliakhovich:2010}.}}
\end{figure}

To leading order, fluctuations in the velocity and density fields are related through the continuity equation which connects the velocity divergence to the time derivative of the density $\dot \delta_c = -\nabla {\bf v_c} $ and $\dot \delta_b = -\nabla {\bf v_b} $ (or in Fourier space $\dot\delta_{k,j} = -ikv_{k,j}$ where $j$ is either $c$ or $b$). This relation ensures that the local values of the  velocity and density fields  are uncorrelated. Moreover, the continuity equation adds an extra factor of $1/k$ to the velocity with respect to the density (where $k$ is the wavenumber), thus suppressing perturbations in velocity on small scales where k is large (and boosting them up at large scales where k is small) with respect to perturbations in density. As a result,  the velocity field is coherent on larger scales than the  density field. This argument is valid for  the velocities of both baryons and dark matter and thus is applicable to their difference (${\bf\vbc}$) as well. Specifically, the lack of power on  small scales, corresponding to  wavenumbers k $>0.5$ Mpc$^{-1}$, observed in the dimensionless power spectrum of the magnitude of the relative velocities, $\vbc$ (see figure \ref{fig:vbc}), points out that the relative velocity field is almost constant on scales smaller than a few comoving  Mpc. It appears that a good approximation for the ``critical scale''  at which the velocities become coherent  is about 3 Mpc \cite{Tseliakhovich:2010}. On much larger scales than the coherence scale, which were out of causal horizon when the relative velocities were generated, fluctuations in $\vbc$ are uncorrelated.  On intermediate scales over the range $k \sim 0.01-0.5$ Mpc$^{-1}$, the power in relative velocity fluctuations oscillate, showing that there is a characteristic scale imprinted in the velocity field, which is the length of the sound horizon at recombination ($\sim 150$ Mpc). These oscillations, seen in the power spectrum of $\vbc$, are set by exactly the same physics as the BAO in the matter power spectrum.

After the CMB decouples,  the sound speed of baryons drops significantly  from $\sim c/\sqrt{3}\sim 1.7\times10^5$ km sec$^{-1}$ to $\sim 6$ km sec$^{-1}$. The relative velocities, that have the root mean square of $\sigma_{bc} \sim 30$ km sec$^{-1}$, become highly supersonic at this stage \cite{Tseliakhovich:2010}.  Since after hydrogen recombination baryons are no longer supported by radiative pressure, they  fall into potential wells formed by dark matter which moves with supersonic speed through the gas. In the absence of any supporting mechanism, the relative velocities, being vector perturbations, decay with time scaling as $(1+z)$, and therefore are mostly important at high redshifts.

\begin{figure}[t]
{\center
\includegraphics[width=2.7in]{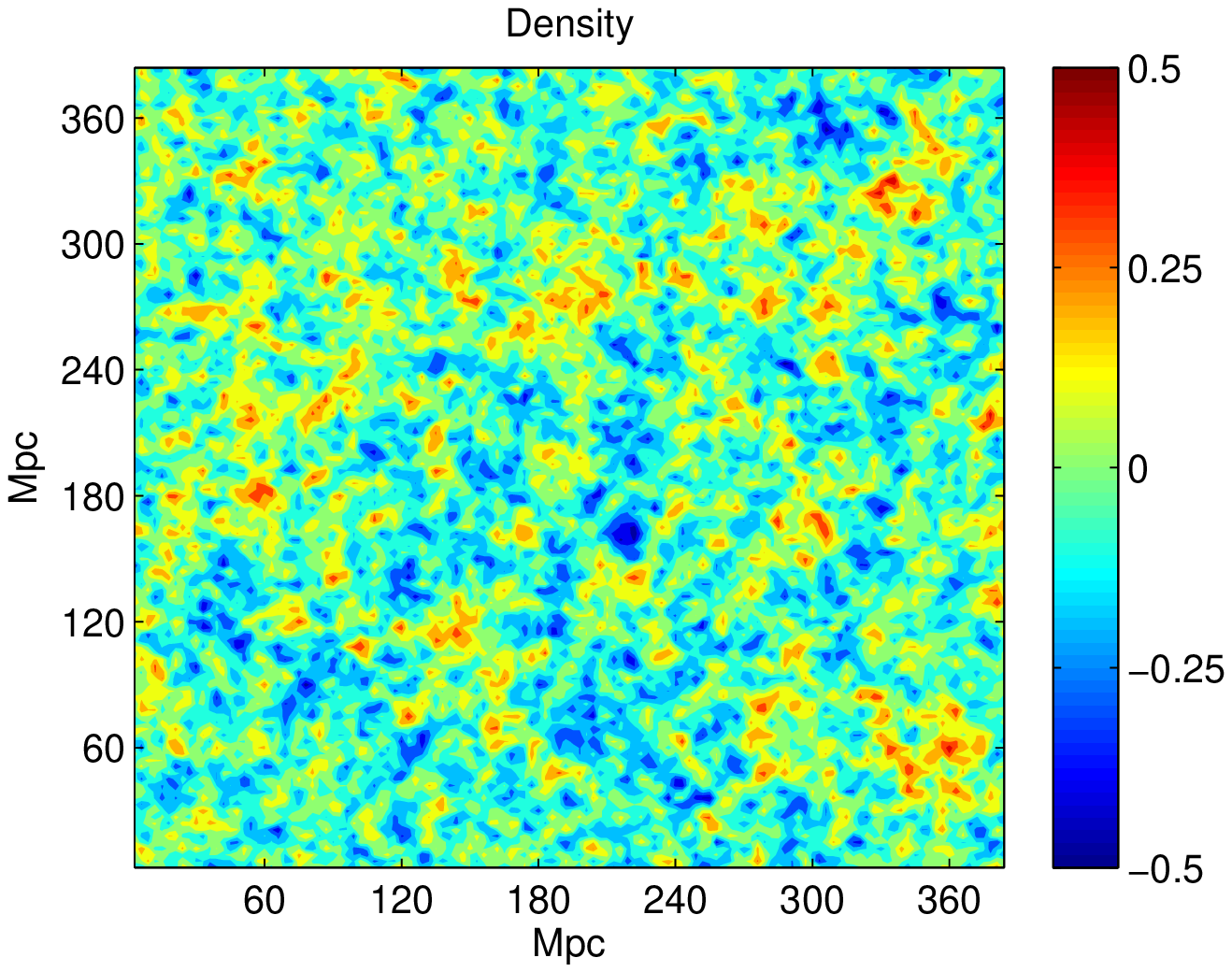}\includegraphics[width=2.7in]{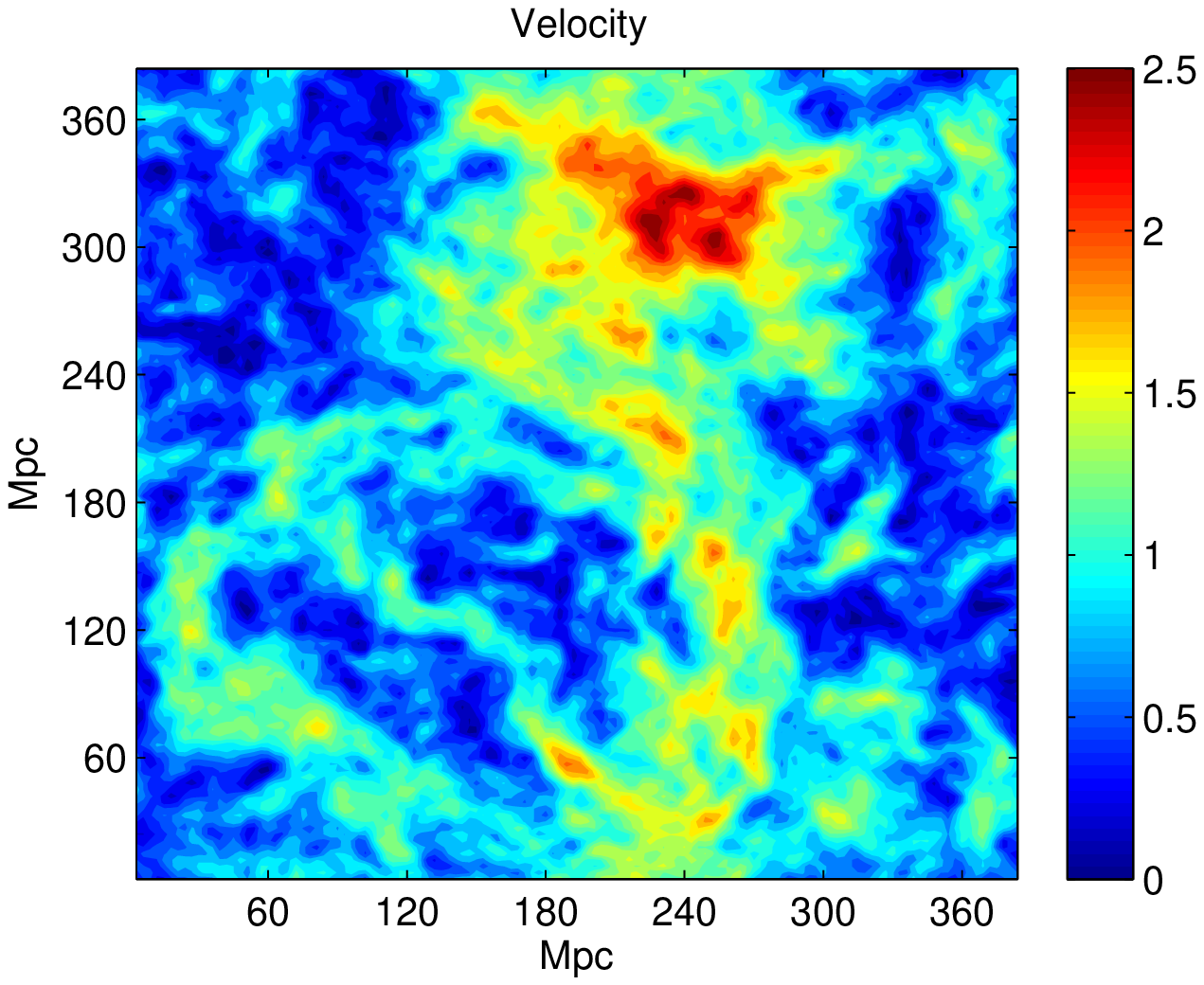}
\caption{\label{fig:IC} Two-dimensional slices  of the relative fluctuation in density at redshift $z = 20$ ({\bf left panel}) and of the  magnitude of relative velocities between baryons and dark matter shown in units of its root-mean-square  ({\bf right panel}) simulated by A. Fialkov et al. (2013) \cite{Fialkov:2013}.}}
\end{figure}

Set by random density fluctuations in baryons and dark matter, the magnitude and direction of ${\bf \vbc}$  randomly varies in space.  Within the assumption of Gaussian initial conditions, the magnitude of the velocity field (three-dimensional Gaussian field) at any scale, i.e. any wavenumber k, follows the Maxwell-Boltzmann distribution function:
\be \label{MBdist}
P_{MB}(v) = \left(\frac{3} {2\pi \sigma^2}\right)^{3/2}4\pi  v^2 \exp\left(-\frac{3 v^2}{2\sigma^2}\right),
\ee
where $\sigma^2$ is the power in the velocity field at the specific scale. Regions with different $\vbc$ are distributed in space according to the Maxwell-Boltzmann distribution (\ref{MBdist}). In fact, the same is valid for any superposition of different modes,  and in particular for a   velocity field smoothed on an arbitrary scale, in which case $\sigma$ should refer to the root-mean-square velocity calculated from the smoothed field. For a smoothing scale around the coherence scale  the root-mean-square velocity is  $\sigma \sim \sigma_{bc}$, which is in accordance with the fact that there is no power on scales smaller than the coherence scale. For such a smoothing of the velocity field,  60$\%$  of space displays velocities $\vbc>1\sigma_{bc}$ and only $6\%$   $\vbc>2\sigma_{bc}$. As a result, the most common patches are those in which the value of $\vbc$ is around $ \sigma_{bc}$. In figure \ref{fig:IC} we show an example of a realization of random  density $\delta_m  $ and velocity $\vbc$ fields smoothed on the scale of 3 Mpc \cite{Fialkov:2013}, where $\delta_m =\left( \Omega_c\delta_c+\Omega_b\delta_b\right)/\left(\Omega_c+\Omega_b\right)$ is the overdensity in matter including both cold dark matter and baryons, while $\Omega_c$ and $\Omega_b$ are the cold dark matter and baryon density parameters respectively \cite{Ade:2013}.   

\section{Impact of the Supersonic Motion on Structure Formation}
\label{Sec:structure}

The process of formation of stars and galaxies begins after the decoupling of the  CMB photons when baryons are no longer supported by pressure and can fall into dark matter potential wells. Such mechanism  has been extensively studied in the absence of relative supersonic motion by means of standard tools to handle the highly non-linear evolution. These tools  are the (1) spherical collapse model, which describes  evolution of a single spherically-symmetric top-hat overdensity,  and (2) the Press-Shechter theory and its variations, which statistically takes into account the abundance of collapsed halos at every instant. The spherical collapse model analytically solves the non-linear problem of gravitational collapse. This model implies  two critical points in the evolution of a structure: the moment of turn-around, when a growing perturbation decouples from the expanding background and  starts to collapse, and  the moment when the non-linear overdensity diverges, which corresponds to the instant at which the linear overdensity ($\delta$ calculated via linear theory) reaches the critical value of $\delta_{crit}\sim 1.686(1+z)$ and the  collapsing halo reaches a state of virial equilibrium\footnote{Note that the numerical factor $1.686$ is changing in time when relaxing the Eistein-de Sitter universe assumption (e.g., S. Naoz, S. Noter, R. Barkana (2006) \cite{Naoz:2006}).  }. Applying the virial theorem one can calculate the radius, mass, and circular velocity of the newly formed halo \cite{Barkana:2001}. The Press-Schechter theory \cite{Press:1974}, based on the Gaussian nature of the initial conditions for structure formation, linear growth and spherical gravitational collapse, is a framework within which the statistical properties of a population of halos at every redshift can be analyzed. Within the Press-Schechter  theory, the number density of halos between M and M + dM is given by
\begin{equation}
\label{eq:dndm}
 \frac{dn}{dM} = \frac{\bar\rho_0}{M}\left|\frac{dS}{dM}\right|f\left(\delta_{crit}(z),S\right),
\end{equation}
where $dn/dM$ is the comoving abundance of halos of mass $M$,  $S(M,z)$ is the variance of matter fluctuations averaged on the mass scale $M$, and  
\begin{displaymath}
f\left(\delta_{crit}(z),S\right) = \frac{1}{\sqrt{2\pi}}\frac{\nu}{S}e^{-\nu^2/2}
\end{displaymath}
 is the mass fraction with $\nu = \delta_{crit}(z)/\sqrt{S}$. However, the Press-Schechter halo abundance fits numerical simulations only roughly and, to achieve better fits, modified approaches are used, such as the one introduced by R. K. Sheth and G. Tormen \cite{Sheth:1999} or by R. Barkana and A. Loeb \cite{Barkana:2004}  who take into account  the bias on structure formation from large-scale density modes. However, this established picture of structure formation appears to be not fully correct when the supersonic motion is accounted for. In particular (as we discuss in detail in section \ref{Subsec:Halos}), the relative velocities affect the statistics of collapsed object in an inhomogeneous way around the Universe.
 
  To correctly predict the halo abundance, one first needs to understand the growth of perturbations when ${\bf \vbc}$ are accounted for. As was shown by D. Tseliakhovich and C. M. Hirata (2010) for the first time, the usual approach which involves linear theory \cite{Barkana:2001} does not always hold even when perturbations are expected to be small. In particular,  it fails to describe structure formation at high redshifts when scales smaller than several Mpc are considered \cite{Tseliakhovich:2010} even when $\delta_m<<1$. Due to the effect of ${\bf \vbc}$, the non-linear terms that are normally ignored in linearized fluid equations because they are small compared to the leading order terms, become non-perturbative when the velocities are treated correctly. To properly describe the growth of structure in baryons and dark matter when the two fluids move supersonically fast one relatively to the other, the complete set of nonlinear fluid equations should be solved. Luckily, due to the coherency  of the streaming velocities on scales smaller than few Mpc, the non-linear equations become  effectively linear at such  scales. Due to this feature,  it is as easy to study the growth of structure on scales smaller than few comoving Mpc with  ${\bf \vbc}$ as in the no-${\bf \vbc}$ case. However in the former case a modified set of equations and a modified Press-Schechter theory should be used.

When interested in scales smaller than the coherence scale of the relative velocity field (and choosing to work in the rest frame of the baryonic fluid, i.e. the frame, in which there is no motion of baryons on large-scale), the velocities of dark matter and baryons can be written as ${\bf \vc} = -{\bf \vbg}+{\bf u_c}$ and  ${\bf \vb} = {\bf u_{b}}$, where ${\bf \vbg}$  is the large scale mode representing  the velocity field smoothed on the coherence scale and satisfying  ${\bf \vbg}\sim{\bf \vbc} $. The large scale mode is constant within the coherence patch and has an amplitude that  decays with redshift $\vbg\propto (1+z)$; ${\bf u_c}$ and ${\bf u_b}$ are, instead, the peculiar  velocities in addition to the large scale mode, representing  the velocity field that has not been smoothed out, and whose evolution is affected by pressure and gravity. Since the relative motion between gas and dark matter is  coherent, the peculiar velocities are expected to be very small. Note that since $u_c$ and $u_b$ are much smaller than $\vbg$, in the rest of this review we will assume ${\bf \vbc} = {\bf \vbg}$, which is correct to leading order in  perturbations.  Now, we can go back to eqs. (\ref{eq:sys1}) and plug in the velocities. Note that the real (small) perturbations are $\delta_b$,  $\delta_c$, ${\bf u_c}$, ${\bf u_b}$ and $\Phi$, while the magnitude of the  background mode can be large, as we argued above. Keeping only the first order terms in the real perturbations and keeping in mind that $\vbg$ is constant, we arrive to the following system of linear equations \cite{Tseliakhovich:2010}: 
\begin{equation}
\begin{array}{l}
\displaystyle\frac{\partial\delta_c}{\partial t}= \frac{i}{a}{\bf \vbc}\cdot {\bf k}\delta_c-\theta_c,\\\\
\displaystyle\frac{\partial\delta_b}{\partial t} = -\theta_b,\\\\
\displaystyle \frac{\partial { \theta}_c}{\partial t} = \frac{i}{a}{\bf \vbc}\cdot {\bf k}\theta_c-\frac{3H^2}{2}
(\Omega_c\delta_c+\Omega_b\delta_b)-2H\theta_c,\\\\
\displaystyle\frac{\partial { \theta}_b}{\partial t}=-\frac{3H^2}{2}(\Omega_c\delta_c+\Omega_b\delta_b)-2H\theta_b+\frac{c_s^2k^2}{a^2}\delta_b,
\end{array}
\label{eq:sys2}
\end{equation} 
where $\theta_i = a^{-1}{\bf \nabla}{\bf v_i}$ is the velocity divergence in comoving coordinates. These are the modified equations, linear in terms of the real perturbations,  properly describing the growth of structure when the relative velocities of baryons and dark matter are taken into account.

\subsection{Dark matter halos and their gas content}
\label{Subsec:Halos}

The Press-Shechter theory and its modifications predict number density of halos at each mass and at every epoch, an essential information for the understanding of the early Universe, since halo abundance is a crucial component which affects the star formation history and thus the amount and properties of stars, black holes, and radiative backgrounds emitted by stars and their remnants. 

\begin{figure}[t]
{\center
\includegraphics[width=3.4in]{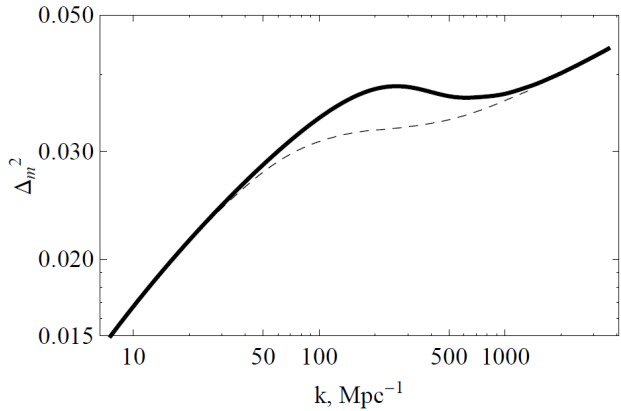}
\caption{\label{fig:PS}Linear power spectrum of matter distribution shown at redshift $z = 40$ without (solid line) and with (dashed line) the effect of ${\bf \vbc}$  included. The figure is adopted from D. Tseliakhovich and C. M. Hirata (2010) \cite{Tseliakhovich:2010}.}}
\end{figure}

The presence of relative supersonic velocities has the potential to dramatically change the landscape of the early Universe, affecting halo abundances and star formation in an inhomogeneous way. To understand the impact of ${\bf \vbc}$ on halo abundance, we first need to solve the system of eqs. (\ref{eq:sys2}) for all possible values and orientations of ${\bf \vbc}$, and  find  the  variance of matter fluctuations $S(M,z,\vbc)$, which now depends on the local value of  $\vbc$, averaged over the orientation of the relative velocities. The total matter power spectrum (including both  baryons and  cold dark matter)  is suppressed by the relative motion \cite{Tseliakhovich:2010} as can be seen from figure \ref{fig:PS} on which dimensionless power spectra $\Delta_m^2(k) = k^3P_m(k)/2\pi^2$ with and without ${\bf \vbc}$ are compared. For instance, at redshift $z = 40$ an average suppression of   $\sim 10\%$ is found on scales $50-500$ Mpc$^{-1}$, and the suppression is maximal ($\sim 15\%$) around Jeans scale $k_J \sim 200$ Mpc$^{-1}$. The range of affected scales is easy to explain: the Jeans scale is exactly the scale at which the evolution of baryonic perturbations is sensitive to both the gravitational attraction  and the total baryonic pressure, thus adding a supplementary pressure term (i. e., the effective anisotropic pressure component due to ${\bf \vbc}$) has a strong impact on the growth of structure at such scales. On the contrary, perturbations on much larger scales would be unaffected by pressure and would  collapse gravitationally, whereas the pressure at smaller-scale prevents the perturbations from collapsing.   Because ${\bf \vbc}$  becomes less important at lower redshifts, the suppression in halo abundance is predicted to be negligible at $z\gtrsim10$.

Applying the modified statistics described by  $S(M,z,\vbc)$ in  eq. \ref{eq:dndm},  one can estimate the effect of the velocities on halo abundance. This was done by means of both seminumerical techniques \cite{Tseliakhovich:2010,  Fialkov:2012, Bovy:2013, Tseliakhovich:2011, Tanaka:2013, Tanaka:2013b} and three-dimensional hydrodynamical simulations \cite{Maio:2011, OLeary:2012,  Naoz:2012}. The seminumerical methods \cite{Tseliakhovich:2010} found the suppression in the  variance of matter fluctuations due to ${\bf \vbc}$  to be  more than 60$\%$ relatively to the case where the large-scale coherent velocities were ignored for halos of M $\sim 10^6$ M$_\odot$ at redshift z = 40, with the strongest effect on halo masses of $10^{6.3}$ M$_\odot$. This is consistent with numerical simulations \cite{Naoz:2012}, which report suppression for halos of masses  M = $5\times 10^4-5\times 10^5$ M$_\odot$ to be $\sim 20 \%$ for the value of $\vbc = \sigma_{bc}$ and $\sim 55\%$ for $\vbc = 3.4\sigma_{bc}$ when evaluated at $z = 25$.  

\begin{figure}[t]
{\center
\includegraphics[width=3.4in]{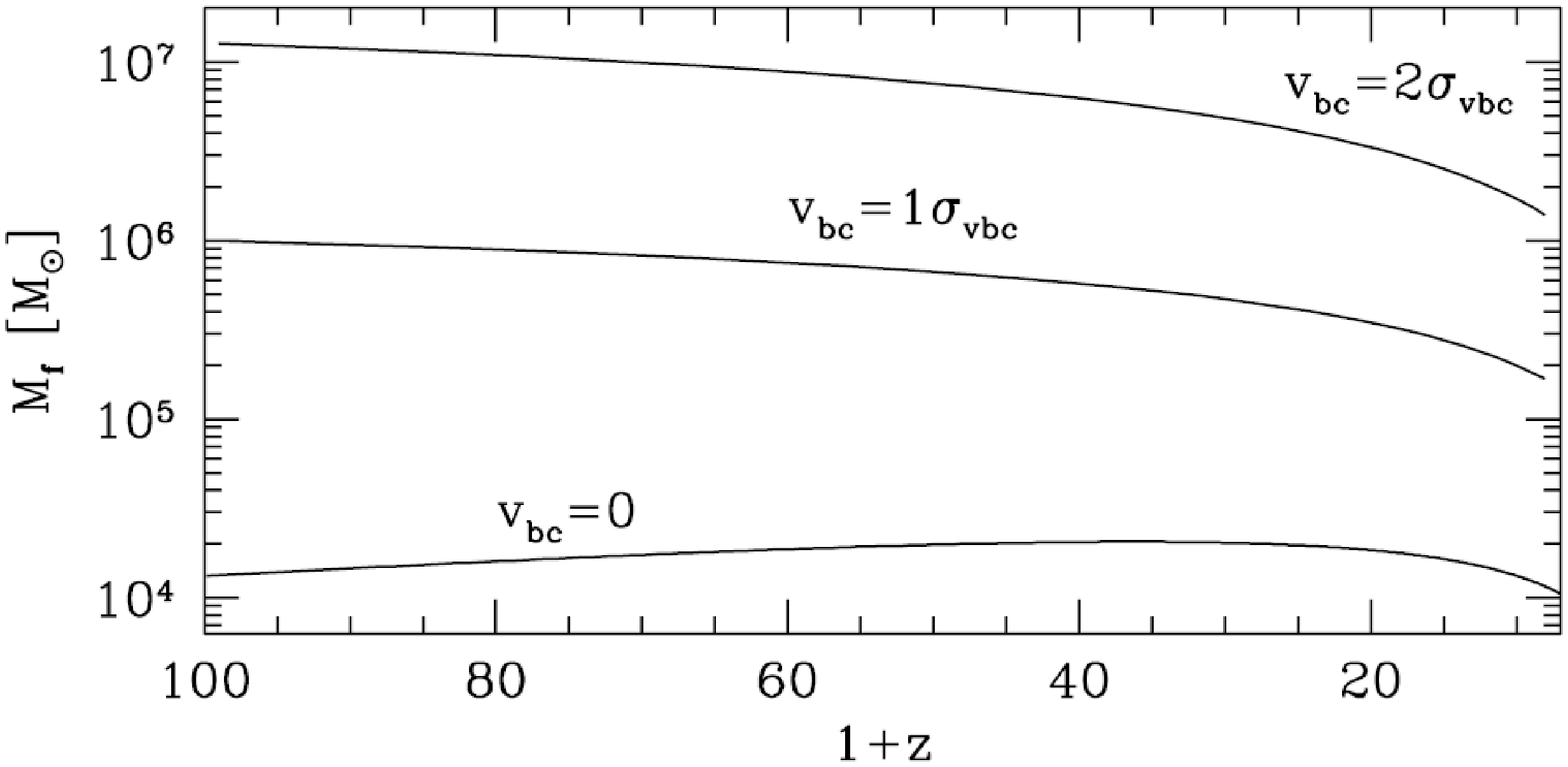}
\caption{\label{fig:Mf}The filtering mass as a function of redshift in the
regions with $\vbc  = 0$, $1\sigma_{bc}$, and $2\sigma_{bc}$. Adopted from S. Naoz, N. Yoshida and N. Y. Gnedin (2013) \cite{Naoz:2013}.}}
\end{figure}

In addition to the suppression in halo number counts, the supersonic relative motion also affects the gas content of  halos of $10^5-10^7$ M$_\odot$ \cite{  Maio:2011, Greif:2011, OLeary:2012, Fialkov:2012, Naoz:2012, Naoz:2013,  Tseliakhovich:2011, Dalal:2010, Richardson:2013}.  Obviously, it gets harder for the dark matter halos to attract baryons which overshoot the halo centers at a supersonic speed. To analyze how much of gas can actually reside in a halo of mass M, we need to compare gravity (which dominates at large scales) and pressure (which dominates at small scales). The relevant mass scale when talking about accretion of baryons by a dark matter halo is believed to be  the {\it filtering scale} M$_F$ \cite{ Naoz:2013, Naoz:2005, Gnedin:1998, Naoz:2007, Naoz:2011}, which approximately is a time-averaged Jeans mass and is identical to the Jeans scale only in a non-evolving background. Because the filtering mass has a memory of the baryonic pressure effects integrated over the entire history of the Universe, whereas the Jeans scale only deals with evolution of perturbations at a given instant in time, it is believed that the filtering scale is more relevant when discussing the growth of gas content of a halo. Although there is no reason to apply the filtering scale derived from linear theory to the non-linear evolution, it was suggested earlier on \cite{Gnedin:2000}  that the filtering mass also describes the largest halo mass whose gas content is significantly suppressed compared to the mean cosmic baryon fraction, thus distinguishing between gas-rich and gas-poor halos. Note that the effect of $\vbc$ was not included in this work. Taking into account the spatially varying speed of sound \cite{Naoz:2005} (but still omitting the effect of $\vbc$), it was shown \cite{Naoz:2007} in agreement with numerical simulations \cite{Naoz:2009, Naoz:2011} that the filtering mass is, in fact, smaller by about an order of magnitude than reported earlier \cite{Gnedin:2000} and gas rich halos can  be present already for $2\times 10^4$ M$_\odot$. However, due to the stream velocity  the baryons overshot the dark matter halo, which results in much higher limit for the minimal mass of gas rich halos \cite{ Naoz:2013, Tseliakhovich:2011} (the dependence of $M_F$ on $\vbc$ and redshift is shown on figure \ref{fig:Mf}). Recent  works show that the  filtering scale computed in linear theory is an accurate match to the actual nonlinear characteristic mass $M_c$ as measured from  simulations, at all redshifts and for all values of the stream velocity considered \cite{Naoz:2013}. By fitting to numerical simulations which include $\vbc$, we can find that the  gas fraction in halos of mass M is given by \cite{Naoz:2013}
\begin{equation}
f_g = f_{b,0}\left[1+\left(2^\gamma-1\right)\left(\frac{M_c}{M}\right)^\beta\right], 
\end{equation} 
where $f_{b,0}$ is the gas fraction in the heavy halo limit, and $\gamma$ and $\beta$ are parameters that depend on $\vbc$ and redshift \cite{Naoz:2013}. In the absence of $\vbc$, $\gamma = 3\beta\sim0.7$. As was noted before, the filtering scale depends on the local value of the relative velocities, and so does the amount of gas in halos.  Since ${\bf \vbc}$  acts as an effective non-isotropic pressure term, it increases the filtering mass; in particular it has been shown \cite{ Naoz:2013, Tseliakhovich:2011} that the effect of streaming motion averaged over the volume of the observable Universe leads to a growth of the filtering mass by an order of magnitude at any redshift in the range $10 < z < 100$. As a result, some of the halos which were thought to be gas-rich in a universe in which the effect of $\vbc$ is neglected, are in reality gas-poor. For instance, as was shown in numerical simulations \cite{Naoz:2012}, at redshift $z = 19$  more than $20\%$ of the halos below 10$^6$  M$_\odot$ in regions with $\vbc = 1.7\sigma_{bc}$ (and $100\%$ of them in regions  with $\vbc = 3.4\sigma_{bc}$) have gas fraction below half of the cosmic mean baryonic fraction.  The semi-numerical analysis carried on by A. Fialkov et al. (2012) \cite{Fialkov:2012} is consistent with the above  numerical results. For instance it shows that at $z = 20$ the streaming motion suppresses the gas content in $10^5$ M$_\odot$ halos by a factor of 2 in average over the observable Universe. The suppression in heavier halos is less significant: the gas content of  $10^6$ M$_\odot$ halos is suppressed by a factor of 1.12, while the deficit in gas content of $10^7$ M$_\odot$ halos is only $\sim 2\%$.
 
\begin{figure}[t]
{\center
\includegraphics[width=4.5in]{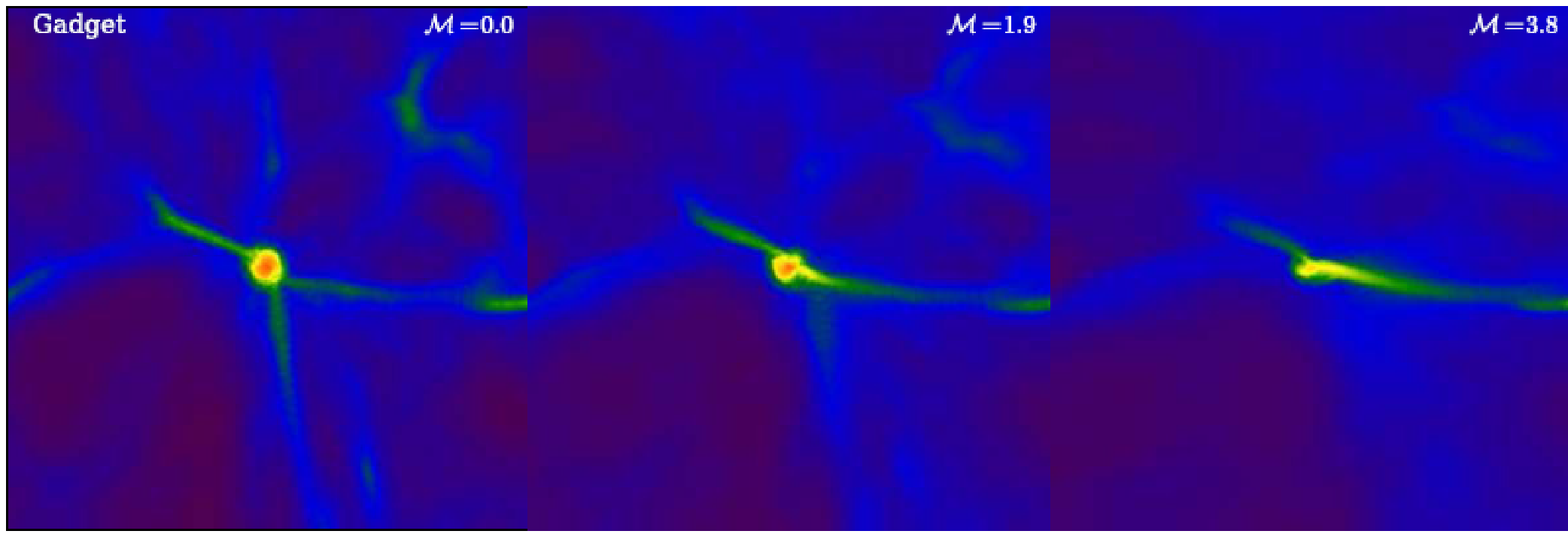}
\includegraphics[width=4.5in]{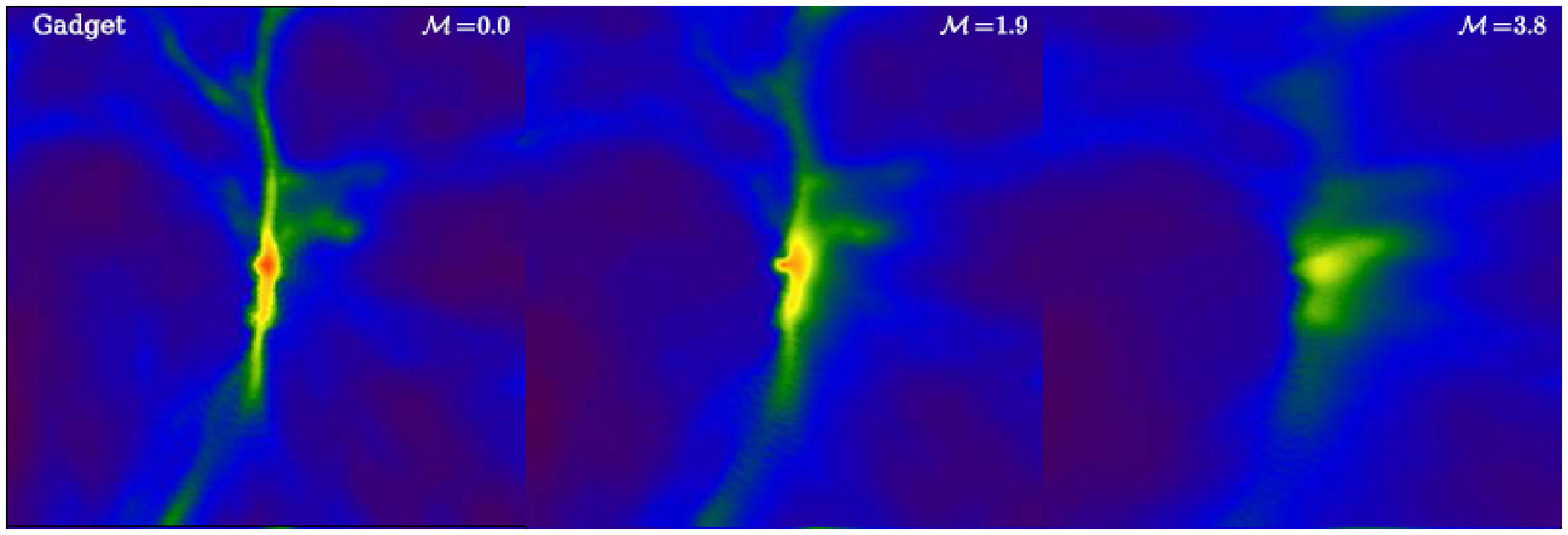}
\caption{\label{fig:Mach} Effect of relative velocities on two isolated  halos at redshift $z = 20$. Plotted at the {\bf top (bottom)} row is a slice (of width 50 kpc h$^{−1}$) through a halo of mass  $2\times 10^6$ M$_\odot$ ( $8\times 10^5$ M$_\odot$). From {\bf left to right} the cases with  $\vbc = 0$,  $\vbc = \sigma_{bc}$,  $\vbc = 2\sigma_{bc}$  are shown with the baryons moving to the right.  The colors indicate the gas density in the range $5\times 10^{−27}$ (blue) - $5\times 10^{−23}$ (red) gr cm$^{-3}$. The figure is adopted from R. M. O'Leary and M. McQuinn (2012) \cite{OLeary:2012}.}}
\end{figure}

Interestingly enough, the temperature profile and gas density inside the halo is also affected by the supersonic relative velocities \cite{Stacy:2011, Greif:2011, OLeary:2012, Richardson:2013}. Recent numerical simulations by  R. M.  O'Leary and M. McQuinn (2012) \cite{OLeary:2012} investigate the way  the inner structure of halos and distribution of gas in halos are modified by ${\bf \vbc}$. Their results show that the central density of  gas is decreased by a factor of two for halos of M$\sim 10^6$ M$_\odot$ in regions with $\vbc = \sigma_{bc}$ and by a factor of five for the same halos in regions with $\vbc = 2\sigma_{bc}$ relatively to the $\vbc = 0$ case. Moreover, the authors find that the orientation of filaments along which the gas flows into halo with respect to the direction of ${\bf \vbc}$ plays an important role and strongly affects the total amount of the accreted gas. In particular, the accretion of gas along filaments which are perpendicular to the direction of ${\bf \vbc}$ suffers the strongest suppression.  Since ${\bf \vbc}$ has random direction, this effect introduces significant additional stochasticity in the baryonic mass fraction of halos at a fixed  mass. Probably, one of the most spectacular effects of ${\bf \vbc}$ is the development of Mach cones generated by dynamical friction in the case of  halos of M$\sim 10^5$ M$_\odot$ in regions with $\vbc\gtrsim 2\sigma_{bc}$, as shown in figure \ref{fig:Mach} adopted from the paper by R. M. O'Leary and M. McQuinn (2012) \cite{OLeary:2012}. 

In addition to the development of Mach cones,  numerous other effects of ${\bf \vbc}$ on the morphology and small-scale structure of individual halos were discussed in the paper by R. M. O'Leary and M. McQuinn (2012) \cite{OLeary:2012}. First, generation of shocks was considered, a feature which could lead to heating of  the gas at high redshifts. However their analysis showed no significant heating of the intergalactic medium. The shocks were, in fact, shown to be largely confined to the immediate region around the halos, especially when $\vbc\neq 0$. Second,  dynamical friction due to the supersonic motion of baryons was shown to  cause some back-reaction onto dark matter halos. The velocity of dark matter gets a kick and the halo starts moving in the direction opposite to the decelerating baryons. However, because typically there  is $\sim 8$ times more mass in the dark matter component, the change in velocity of dark matter halos is negligible compared to the magnitude of ${\bf \vbc}$. Third, it was shown that relative velocities can generate vorticity thus seeding turbulence and magnetic fields \cite{OLeary:2012, Pudritz:1989}, which in turn may alter the properties of first stars. The primary way to generate cosmological vorticity is via curved shocked fronts, and ${\bf \vbc}$ clearly alters the morphology of shocking regions. However, the authors found only  a modest difference in the voriticity magnitude  with $\vbc \neq 0$ in halos of $\sim 10^6$ M$_\odot$ in their simulations at z = 20, when most of the gas had already decelerated. Fourth, it was shown that  when there is a nonzero velocity difference between  dark matter and baryons, their profiles no longer trace each other at small scales even in the most massive halos with masses of $\sim  10^6$ M$_\odot$. The authors in fact found that the gas density peak is normally located downwind from the dark matter density peak and is shown to be typically off-set by as much as $\sim r_{200}$ (where $r_{200}$ is defined as the distance from the halo density peak at which the dark matter overdensity falls below 200). In addition to that, the gas profile is shallower when velocities are significant, and the maximum gas density can be suppressed on average by roughly an order of magnitude between the cases $\vbc = 0$ and $\vbc\sim 2\sigma_{bc}$ for halos with masses of $\sim 10^4$ M$_\odot$.  We refer the interested reader to the original paper by  R. M. O'Leary and M. McQuinn (2012) \cite{OLeary:2012} for a much more complete and detailed discussion. 

The fact that the  profiles of baryons and dark matter no longer trace each other at small scales when ${\bf \vbc}$ is present was further explored  in a recent work by S. Naoz and R. Narayan (2014) \cite{Naoz:2014}. The authors confirm  that when the perturbations become non-linear and collapse, the baryon clump forms at a different spatial location than its dark matter counterpart. It is shown that the separation between the peaks of $\delta_b$ and $\delta_c$ exceeds the virial radius of the dark matter  halo for structures of baryon masses within $\sim 10^5 < $ M$_b <$ few$\times 10^6$ M$_\odot$ for $\vbc = \sigma_{bc}$. In this setup, tidal forces from other nearby objects may be capable of unbinding the baryons from their dark matter counterpart, leading to the formation of a dark matter-free baryonic clump. This scenario provides a possible explanation for how the observed globular clusters, which contain practically no gravitationally bound dark matter \cite{Heggie:1996,  Bradford:2011, Conroy:2011, Ibata:2013},  formed. Remarkably, the predicted mass range of the long-lived baryonic clumps separated from their dark matter halos ($\sim 10^5 < $ M$_b <$ few$\times 10^6$ M$_\odot$ for $\vbc = \sigma_{bc}$) agrees well with the masses of the observed present-day globular clusters. Finally,  the corresponding gas-poor dark matter halos may be related to the present day dark satellites or ultra-faint galaxies, thus alleviating the missing satellite problem.

\subsection{Formation of the first stars}
\label{Sec:FS}
In the previous section, we have considered two distinct effects of relative velocities on structure formation, namely  the suppression of halo abundance and the suppression of the gas content of each individual halo.  As we advocated above, a halo with a mass greater than the filtering mass  is capable of  accreting gas efficiently. However, a totally different  question is whether or not such a gas-rich halo is capable of forming stars out of the accumulated gas. This issue was first tackled by using  numerical simulations \cite{Maio:2011, Stacy:2011, Greif:2011, OLeary:2012} and the results are reviewed in this section. 

In order to form stars inside a specific halo, the halo should be massive enough to gravitationally accelerate the gas to high enough velocities so that the gas can, after an initial heating stage,  efficiently  cool by radiative processes, and finally condense  into stars. The minimal mass of a halo that can complete this cooling
process  is referred  to as the {\it minimal cooling mass}, $M_{cool}$. Alternatively, the same threshold can also be described by the  minimum circular velocity needed for efficient cooling, $V_{cool}$, which for a halo of mass $M$ and virial radius $R$ is related to the minimal cooling mass via the standard relation $V_{cool} = \sqrt{GM_{cool}/R}$, where $G$ is Newton's constant. First stars are thought to be formed out of molecular hydrogen which is the lowest temperature coolant available in the metal-free primordial gas, since it only needs to be heated up to $T \ge 300$ K to initiate the radiative cooling process. For comparison, the next available coolant is atomic hydrogen, which radiatively cools when it reaches a much higher temperature of $T\sim 10^4$ K. Since the cooling temperature of molecular hydrogen is so low, stars can form even in relatively light halos of mass $\sim 10^5$ M$_\odot$ \cite{Haiman:1996, Tegmark:1997, Machacek:2001,Abel:2002} (while in the case of atomic hydrogen the mass rises to $\sim 10^7$ M$_\odot$).

The presence of streaming velocities is expected to wash out baryons from dark matter halos and to hinder their collapse, as was previously noted in  section \ref{Subsec:Halos}. Therefore, when $\vbc\neq 0$ one can anticipate that a heavier halo is needed to accelerate and heat up the gas to the temperature at which the cooling becomes possible. To quantify the effect of ${\bf \vbc}$ on cooling and star formation, which are highly non-linear processes, numerical simulations are essential as has been done in recent works \cite{Maio:2011, Stacy:2011, Greif:2011, OLeary:2012}. The first hydrodynamical numerical simulation in which the effect of ${\bf \vbc}$ was closely  examined was performed by U. Maio, L. V. E. Koopmans and B. Ciardi (2011) \cite{Maio:2011}, who used Smooth Particle Hydrodynamics (SPH code) to follow $320^3$ particles in gas and dark matter within a 1 Mpc box, and reported  a reduction in  star formation rates, halo abundance and gas fraction in halos. However in this work no noticeable  effect of ${\bf \vbc}$ on the minimum cooling mass was reported. 

\begin{figure}[t]
{\center
\includegraphics[width=3.1in]{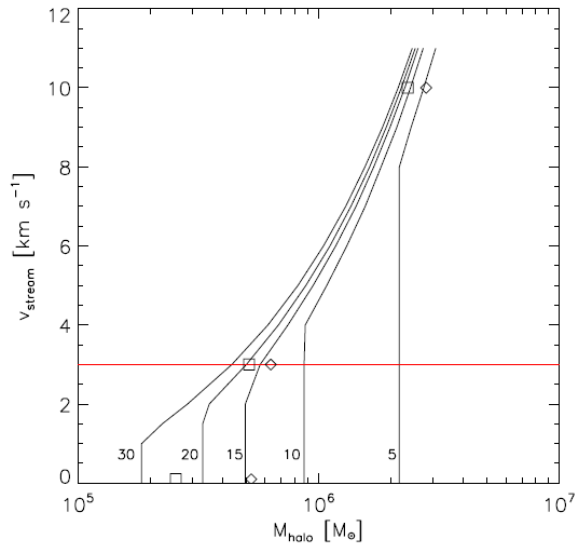}
\caption{\label{fig:Mcool1}Results of an SPH simulation \cite{Stacy:2011} demonstrating the effect of relative streaming on the minimum halo mass into which primordial gas can collapse. Each line represents the necessary halo masses for baryon collapse at a different redshift, marked in the plot. The diamonds represent the final halo masses found in “standard collapse” simulations ($z_{collapse} = 14$ for no streaming), and the squares represent masses from the “early
collapse” simulations ($z_{collapse} = 24$ for no streaming). Note that the halo mass does not noticeably increase unless the initial streaming velocities are very high ($\vbc> \sigma_{bc}$). Also note that halos collapsing at high redshift are more affected by relative streaming, as the physical streaming velocities are higher at these early
times. The figure is adopted from A. Stacy, V. Bromm and A. Loeb (2011) \cite{Stacy:2011}.}}
\end{figure}

An increase of $M_{cool}$  was instead pointed out for the first time  by two groups who independently performed simulations using two different algorithms: an SPH code \cite{Stacy:2011} in which the evolution of $128^3$ particles of each type was followed within a $0.1 h^{-1}$~Mpc box, and  a moving-mesh \cite{Greif:2011} (hereafter MMH) code, which achieved a much higher resolution and followed  $256^3$ particles in a 0.5~Mpc box. To model star formation, both simulations mentioned here tracked the abundance and the cooling of the chemical components that filled the early universe\footnote{The relevant chemical network realized in the simulations \cite{Stacy:2011, Greif:2011} included the evolution of H, H$^+$, H$^-$, H$_2^+$, H$_2$, He, He$^+$, He$^{++}$, e$^-$, D, D$^+$, D$^-$, HD and HD$^+$, which is determined by processes such as H and He collisional ionisation, excitation and recombination cooling, bremsstrahlung, inverse Compton cooling, collisional excitation cooling via H$_2$ and HD, and H$_2$ cooling via collisions with protons and electrons.}, along with the effect of dark matter and gravity.   To be precise, these simulations provide the mass of a halo when it first allows a star to form, i.e., when it first contains a cooling, rapidly-collapsing gas core.  The two simulations \cite{Stacy:2011, Greif:2011} agreed on that relative velocities increase the mass of star forming halos  by roughly $60\%$ in  regions with  significant ${\bf \vbc}$ (an example of an increment in the halo mass with the streaming velocity is shown on figure \ref{fig:Mcool1}). In addition,  star formation was found to be delayed by $\Delta z \sim 5$ in average, a different effect  from the suppression of total amount of gas in halos considered in section (\ref{Subsec:Halos}), which implies a smaller number of stars in a given halo at a given time. Moreover, this suppression is not simply related to the total amount of accreted gas, in fact, even allowing the halo affected by ${\bf \vbc}$ to accrete the same total gas mass of its no velocities counterpart, it will still fail
in forming stars. The modified cooling criterion  is explained by the fact that the internal density and temperature profiles of the gas are modified when $\vbc\neq0$ \cite{Stacy:2011, Greif:2011, OLeary:2012,  Richardson:2013}, as was briefly mentioned in section (\ref{Subsec:Halos}).

\begin{figure}[t]
{\center
\includegraphics[width=3.5in]{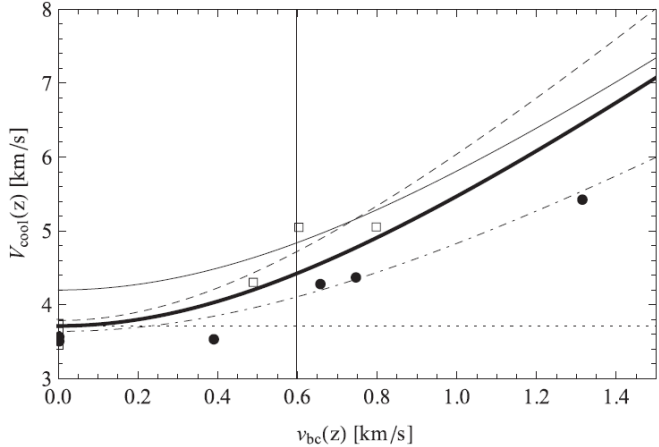}
\caption{\label{fig:vc} Minimum halo circular velocity for gas cooling via molecular hydrogen versus the bulk velocity $\vbc$ at halo virialization. Data points from fig. 2 of A. Stacy, V. Bromm and A. Loeb (2011) \cite{Stacy:2011} are shown with circles and from fig. 3 of T. Greif et al. (2011) \cite{Greif:2011} are shown with squares. Fits to each set of simulations are shown (dot–dashed and dashed, respectively). An ``optimal'' fit to all the data points together (thick solid line), a ``fit'' to AMR simulations (regular solid line, see the reference \cite{Fialkov:2012} for a detailed description) and the case of no streaming velocity (dotted line, based on the optimal fit) are also shown. The vertical solid line marks the root-mean-square value
of $\vbc$ at z = 20. The figure is adopted from A. Fialkov et al (2012) \cite{Fialkov:2012}.}}
\end{figure}

In order to compare the effect just discussed to the suppression in halo abundance and gas content, it is convenient to express the cooling threshold in terms of the halo circular velocity $V_{cool}$. Simulations, in which ${\bf \vbc}$ is ignored \cite{ Abel:2002, Fuller:2000, Bromm:2002,Yoshida:2003a, Yoshida:2003b, Yoshida:2006, Bromm:2004,Reed:2005, Gao:2006,Maio:2010,Maio:2011a,Petkova:2011,Turk:2011,Stacy:2012a, Wise:2007}  find an approximately redshift-independent threshold $V_{cool,0}(z) $ of about  $ 4$ km sec$^{-1}$, which slightly varies from simulation to simulation; this is naturally expected since molecular cooling turns on essentially at a fixed gas temperature, and the halo circular velocity determines the virial temperature to which the gas is heated. Thus, the limit of zero bulk velocity simply gives a fixed threshold $V_{cool,0}(z)$. When relative velocities are added, the minimum circular velocity  may vary as a function of both redshift and $\vbc$. A. Fialkov et al. (2012) \cite{Fialkov:2012} fit the results of simulations \cite{ Stacy:2011, Greif:2011} and  find the following dependence  
\begin{equation}
V_{cool}(z) =\left\{ V_{cool,0}(z) ^2+\left[\alpha \vbc \right]^2 \right\}^{1/2},
\label{eq:vcool}
\end{equation}
where $\alpha$ is a free parameter which sightly varies from simulation to simulation and which expresses how strongly the cooling criteria depend on $\vbc$. Since the dependence of the circular velocity $V_{cool}$ on redshift effectively only appears through its dependence on $\vbc$, it follows that  the star-formation threshold in a patch with a statistically rare, high value of $\vbc$ at low redshift is the same as the threshold in a patch with the same (but now statistically more typical) value of $\vbc$ at high redshift. This should be the case during the era of primordial star formation, before metal enrichment and other feedbacks complicate matters. How well the function fits the data points \cite{Stacy:2011, Greif:2011} is shown in figure \ref{fig:vc}, with  ``optimal parameters'' found by A. Fialkov et al. (2012) given by  $ V_{cool,0} =  3.714$ km s$^{-1}$ and $\alpha = 4.015$. Note that  the $V_{cool,0}-\alpha$ parametrization catches only the average behavior and  does not account for the significant halo-to-halo stochasticity due to the mutual orientation of ${\bf \vbc}$ and filaments along which gas flows into the halo, as found in some numerical simulations \cite{OLeary:2012, McQuinn:2012}, where the values  $\alpha \sim  4 −- 6$ for $V_{cool,0} = 3.7$ km s$^{-1}$ are preferred. It can also be noted that choosing other cooling criteria, e. g. atomic cooling, would mainly change $ V_{cool,0}$, leaving the range for $\alpha$ largely  unmodified \cite{McQuinn:2012}. 

In total, adding ${\bf \vbc}$ increases the circular velocity needed for efficient condensation and star formation, and thus it also increases the minimal cooling mass $M_{cool}$ of halos in which stars can form. In  patches with no relative motion, $M_{cool}$ drops rapidly with redshift (see fig. 1 of A. Fialkov et al. (2012) \cite{Fialkov:2012}); however, in  regions with velocities close to  the root-mean-square value of $\vbc$ the higher bulk velocity at high redshift implies that  higher halo masses are needed for efficient molecular cooling. In particular, at redshift $z = 20$ a patch with $\vbc = 0$ will form stars in halos with masses higher than $\sim 3.6 \times 10^5$ M$_\odot$, while a patch at the root-mean-square value of $\vbc$ will have  a minimum cooling mass of $\sim 6.0 \times 10^5$ M$_\odot$ when the optimal fit from eq. \ref{eq:vcool} is used \cite{Fialkov:2012}. Thus, in patches with low bulk velocity (which are rare) we expect stars to form earlier, since the halos with lower masses are more abundant and form earlier in the hierarchical picture of structure formation.

\subsubsection{The complete effect of ${\bf \vbc}$ on stars}

\begin{figure}[t]
{\center
\includegraphics[width=4in]{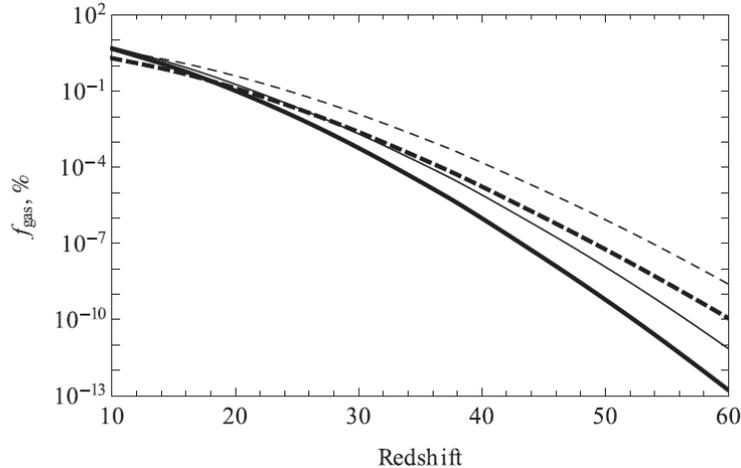}
\caption{\label{fig:fgas} Global mean gas fraction in star-forming halos (solid curves)
and in minihaloes, i.e. halos below the cooling threshold (dashed curves).
The results, based on the optimal fit (equation 3 of A. Fialkov et al. (2012) \cite{Fialkov:2012}), are shown after averaging
over the distribution of relative velocities (thick curves), or in the case of no
relative motion, i.e. for $\vbc$ = 0 (thin curves). The figure is adopted from A. Fialkov et al. (2012) \cite{Fialkov:2012} (and uses the prescription for filtering mass as in D. Tseliakhovich, R. Barkana and C. M. Hirata (2011) \cite{Tseliakhovich:2011}).}}
\end{figure}

Only when we combine together the three effects discussed above: suppression of  halo abundance,  suppression of the gas content and the increment of the minimal cooling mass, we have the complete picture of the impact of ${\bf \vbc}$ on star formation. To get the feeling of how strong  the total effect of ${\bf \vbc}$ might be, we calculate the fraction of gas in two types of halos (1) star-forming halos, and (2) low-mass halos which are too light to form stars (with mass below the minimal cooling threshold).   The first category consists of large halos in which the gas can cool via molecular hydrogen cooling or other cooling mechanisms (e. g.,  atomic or metal line cooling). These halos are presumed to be the sites of formation  for the  first stars and galaxies, and are obviously most important to understand since the stellar radiation they induce is in principle observable, and produces feedback on the intergalactic medium and on other nearby star forming sites. The gas fraction in this type of halos in a patch with the magnitude of relative velocity $\vbc$  is given by 
\be f_{gas}(>M_{cool})=\int_{M_{cool}}^{\infty} \frac{M}{\bar\rho_0} \frac{dn}{dM}\frac{f_g(M)}{f_b}dM\,\ee
where $\rho_0$ is the background density, and $M_{cool}$, $dn/dM$ and $f_g$ depend on the value of $\vbc$ and are either calculated as explained in previous sections or produced in numerical simulations. The effect of ${\bf \vbc}$ on $f_{gas}(>M_{cool})$ averaged over all possible values of the relative velocities is shown on figure \ref{fig:fgas}. 

Also interesting, in principle, is the second category of halos: namely the smaller ones in which the gas accumulates to roughly virial density and yet cannot cool. This halos remain star-less and act as reservoirs of metal free gas down to
lower redshifts  than in the case without ${\bf \vbc}$.  There are also empty dark matter halos which cannot accrete gas at all, and contribute to the Universal evolution only gravitationally. The effect on the gas fraction in this type of halos $f_{gas}(<M_{cool})$  is also demonstrated  in figure \ref{fig:fgas}.

Comparing the two categories of halos in figure \ref{fig:fgas}, we find that the relative suppression of the gas fraction due to ${\bf \vbc}$  is larger for the star-less halos than for  the star-forming halos at low redshift \cite{Fialkov:2012}. For instance, at redshift $z=20$ the bulk velocities reduce the mean gas fraction in star-forming halos by a factor of 1.8 and  that in star-less halos by a factor of  3.1. On the other hand, the relative suppression of the star-forming halos increases faster with redshift, and eventually it becomes larger than that of the empty halos (beyond $z \sim 50$). 

To summarize, the presence of relative velocities produces three distinct effects on star formation: (i) suppression of the halo abundance ($dn/dM$), (ii) suppression of the gas content within each halo ($f_g(M)$), and (iii) boosting of the minimal cooling mass (introduced here through $V_{cool}(z)$). It turns out that for star-forming halos, the suppression of gas content is always the least significant effect \cite{Fialkov:2012} (e.g., on its own it leads to suppression by a factor of 1.13 at $z=20$), while the cooling mass boost becomes the  most important effect above $z=28.5$ (on its own it causes a suppression by a factor of 1.26 at $z=20$), and the halo abundance cut is most important at lower redshifts (on its own it suppress star formation by a factor of 1.43 at $z=20$). For the low-mass star-less halos, the boosting of the minimum cooling mass acts as a (small) positive effect, since it moves halos from the star-forming to the star-less category (e.g., boost by a factor of 1.10 on its own at $z=20$), while the other two effects are larger and comparable (e.g., at $z=20$ the suppression of gas content would give a reduction by a factor of 2.17 on its own, and the halo abundance cut would give a suppression factor of 1.74). 

The average delay in star formation, introduced by ${\bf \vbc}$, is most significant at high redshifts and becomes negligible at $z\lesssim 20$. In particular,  the formation of the very first star in the volume of the observable universe \cite{Naoz:2006} is delayed \cite{Fialkov:2012} by  $\Delta z = 5.3$  (i.e. by $\Delta t = 3.6$ Myr)  in consistency with the delay in star formation found in numerical simulations \cite{Greif:2011}. With the relative motion included, the very  first star  is  most likely to form at  $z=64.6$ (corresponding to $t=32.9$~Myr after the Big Bang), instead of $z \sim 70$ as is expected when velocities are neglected. In practice, the delay in star formation is inhomogeneous: the rare patches of the sky with $\vbc\sim 0$ will form stars earlier than the more common patches with $\vbc\sim \sigma_{bc}$, and the  same pattern is inherited by all the products created by stars. For instance, radiative backgrounds produced by stars are expected to be inhomogeneous, and so is the heating of the intergalactic medium. The inhomogeneous radiative backgrounds are discussed in more details in section \ref{Sec:observ}.

\subsubsection{A competing effect: negative feedback to star formation} 
\label{Sec:LW}
Astrophysical phenomena can cause similar effects to that of  ${\bf \vbc}$ in terms of shutting down star formation in light halos. For example,  ultra-violate (UV) photons produced by the first stars, in particular photons in the Lyman-Werner (LW) band with energies  11.2-13.6 eV, can dissociate molecular hydrogen, and, therefore, prevent star formation via H$_2$ cooling \cite{Haiman:1997}. The formation of the first stars via cooling of molecular hydrogen is a highly non-linear process. Although it  can be mimicked by numerical simulations, they tend to oversimplify the potentially fatal effect of the LW background on H$_2$ cooling in light halos. In fact, until recently the effect of LW photons on star formation was only considered in a simplified manner by taking the intensity of the LW radiative background to be constant and equal to its value after halo virializetion \cite{Machacek:2001, Wise:2007, OShea:2008}, which most likely led to an overestimated effect of LW photons. The degree to which the effect of the LW feedback may be overestimated is unknown at the moment and can be parameterize by introducing a delay in its impact on molecular hydrogen \cite{Fialkov:2013}. However, the most recent  numerical estimates \cite{Visbal:2014},  that follow the density, chemical abundances, and temperature of gas in the central regions of dark matter halos, including hierarchical growth and a time evolving LW background, suggest that the  LW background most probably acted with almost no delay.  At each redshift $z$ the authors compare the results obtained with a consistently evolving LW feedback to the results of a similar calculation where the LW flux was set to the value at a previous redshift step and find less than $30\%$ difference in fraction of halos in which gas has cooled by redshift $z$. This result hints on that the LW feedback was most likely very efficient in destroying molecular hydrogen as soon as it built up, and, therefore, star formation probably proceeded in atomic cooling halos, which are too heavy to be affected by ${\bf \vbc}$. In such a scenario we would expect ${\bf \vbc}$ to be important only at the very beginning of star formation.

The negative feedback by LW photons was shown to  boosts the minimal cooling mass, $M_{cool}$, according to \cite{Machacek:2001, Wise:2007, OShea:2008}
\be \label{Mcool0} M_{cool}\left(J_{21}, z\right) = M_{cool,0}(z)\times\left[1+6.96\left(4\pi J_{21}\right)^{0.47}\right], 
\ee where $J_{21}$ is  the Lyman-Werner intensity in  $10^{-21}$erg s$^{-1}$cm$^{-2}$Hz$^{-1}$sr$^{-1}$ units and  $M_{cool,0}(z)$ is the value of the minimum cooling mass in the standard case with no Lyman-Werner background. This result is incomplete since it does not account for the relative velocity ${\bf \vbc}$, which as we have seen also boosts the  minimum cooling mass \cite{Stacy:2011, Greif:2011, Fialkov:2012}. To estimate the joint effects of the relative velocities and the negative feedback in the absence of more adequate simulations, the parameter  $M_{cool,0}(z)$ in eq.~\ref{Mcool0} can be modified \cite{Fialkov:2013} to include the effect of ${\bf \vbc}$ by calibrating it to the value of the minimal cooling mass from section \ref{Sec:FS}, which includes  the relative velocities and redshift effects, but  neglects the LW background.

The fact that the LW background decreases the amount of  H$_2$ gas available for cooling  suggests that this  negative feedback may have  a similar impact  as ${\bf \vbc}$ on structure formation, resulting in a suppression of the amount of gas available for star formation \cite{Fialkov:2013}. If the effect of the feedback is strong, stars do not form via molecular cooling, and halos of masses $\sim 10^5-10^6$ M$_{\odot}$ are empty of stars at $z\gtrsim 20$, in a similar manner to the effect of the relative velocities which wash the gas out of the halos.  However, the timing of the two effects is different: while in regions where $\vbc$ is high, star formation in   $\sim 10^5-10^6$ M$_{\odot}$  halos is hindered form the beginning,  the LW background needs to build up and reach an intensity of at least $J_{21}\sim 10^{-5}$  erg s$^{-1}$cm$^{-2}$Hz$^{-1}$sr$^{-1}$ before having any effect on the H$_2$ abundance.

Another potentially interesting effect which can be degenerated with the effect of $\vbc$ is that of positive feedback to star formation by X-ray photons \cite{Machacek:2003}. X-rays emitted by early sources could catalyze production of molecular hydrogen by increasing the electron fraction \cite{Haiman:1996, Haiman:2000,  Oh:2001, Ricotti:2001}.  However the positive effect of X-rays was found to be very mild \cite{Machacek:2003}.

\subsection{Black Holes}

First stars formed in a metal free universe and were most probably very massive. The initial mass function of first massive stars  is still very uncertain and a significant effort has been put into modeling the first star-forming systems (see a review by V. Bromm (2013) \cite{Bromm:2013} for a detailed discussion). Massive stars in simulations follow two main evolution scenarios at the end of their lives: (1) ending their lives through a direct collapse to a massive black hole if they are non-rotational and have masses in the range $\sim 40-140$ M$_\odot$ or above $\sim 260$ M$_\odot$, or (2) exploding as pair-instability supernovae without leaving a black hole remnant  if their masses are within $\sim 140-260$ M$_\odot$. Stars with moderate initial masses in the range $\sim 25-40$ M$_\odot$ leave black hole remnants but also eject metals to the intergalactic medium, whereas lighter stars  do not leave any black hole remnants. However, recent simulations \cite{Stacy:2011b} find that the first  stars  produced in halos of M$\sim 10^6$ M$_\odot$ at $z\sim 20$ have significant rotation, which can prevent the formation of a  black hole or reduce its mass.  More details on life and death of first massive stars can be found in a paper by A. Heger and S. E. Woosley (2002) \cite{Heger:2002} and in a recent review by Z. Haiman (2013) \cite{Haiman:2013}. Latest  numerical simulations seem to prefer typical  stellar masses  around $\sim 30-140$  M$_\odot$ \cite{Bromm:2013}, value which is still largely unconstrained by observations.

In this section we review the effect of the relative supersonic motion on  black holes left as remnants at the end of the lives of first stars. Since the relative supersonic motion affects formation and the spatial distribution of  early stars, it is also expected to have an impact on the population of black holes. In fact, it can help to resolve a puzzle related to recently observed supermassive black holes of masses around $10^9$ M$_\odot$  found at unexpectedly high redshifts of 
 $z \gtrsim	 6$ \cite{Fan:2006, Mortlock:2011,Venemans:2013}.  The existence of so massive black holes so early in cosmic history is still a mystery and is not easy to explain \cite{Haiman:2013, Volonteri:2010}.  One  hypothesis for their origin proposes  they grew via rapid accretion from the remnants of first generation $40-140$ M$_\odot$ stars, which form at  $z > 20-40$ in halos of M$\sim 10^5-10^6$ M$_\odot$, incidentally is exactly the  regime where baryonic streaming motion most strongly affects star formation. To grow into the observed supermassive black holes, the seed black holes, which are expected to be born with  $\sim 40\%$ of the parent star mass \cite{Zhang:2008}, should accrete at a rate close to the Edington limit from $z\sim 40$ when the seeds are formed to the redshift of observation.   Quite clearly a very hard setup to achieve.

An alternative scenario for the  formation of supermassive black holes is via direct collapse of a cloud of metal-free gas of temperature $T>10^4$ K. The essential detail of this possibility is that the fractions of strong coolants that trigger star formation (e.g. molecular hydrogen, metals and dust) has to be kept minimal for it to work. If this condition is satisfied,  the cloud collapses to form a $\sim 10^5$ M$_\odot$ star-like object that eventually forms a massive black hole with a similar mass. Of course, a halo able to heat gas up to such high temperatures would be massive enough to trigger star formation unless some mechanism, such as the presence of a strong LW background or ${\bf \vbc}$, were to be present and prevent formation of stars as discussed in section \ref{Sec:FS}. There are also  alternative ways to suppress the $H_2$ fraction (and thus formation of first stars) and to allow direct collapse in the absence of the mechanisms we just mentioned, most noticeably  collisional dissociation can be effective if the gas is hot and dense ($T>10^4$ K and $n>10^3$ cm$^{-3}$).  See  review by Z. Haiman (2013) \cite{Haiman:2013} and references therein for more details on this aspects of primordial star formation.  

While, as we have already seen in previous sections, ${\bf \vbc}$ has the capability to hinder star formation, interestingly enough, it can have either a negative or a positive impact on the formation of  supermassive black holes when applied to, respectively, the first and second scenarios presented above. In the first scenario for the origin of black holes, ${\bf \vbc}$ is likely to  have a weak negative effect on formation of the seeds  from first stars \cite{Tanaka:2013}. This claim is supported by a semianalytic study \cite{Tanaka:2013}  which uses Monte Carlo merger trees to simulate the hierarchical assembly of dark matter  halos and the growth of their nuclear black holes \cite{Tanaka:2013}. Their analysis takes into account not only  the   effect of ${\bf \vbc}$ on formation of seed black holes, but also the  important effect that  heating of the inter-galactic medium has on the Jeans mass. Heating increases the Jeans scale by boosting up the sound speed of baryons (eq. \ref{eq:cs2}) so that gas in low-mass halos can no longer collapse. As a result, the number densities of  seed black holes decreases. Importantly, the effect of heating and ${\bf \vbc}$ have different timing: the former becomes important at lower redshifts than the latter since a population of heating sources has to build up first in order to significantly affect the gas. The results of this semianalytic study show that the effect of streaming motion on observable quantities (such as the total mass density of black holes, their mass function, their contribution to the  heating of intergalactic medium, and their merger rates) in a  scenario where black holes grow at 2/3 of the Eddington limit during their lifetime ($\sim 3\times10^7$ yr) is marginal \cite{Tanaka:2013}\footnote{Note, however, that  this analysis does not account for the suppression in the halo abundance due to ${\bf \vbc}$ and thus underestimates the overall  effect of the relative velocities.}. However, suppression of seed  black hole formation by ${\bf \vbc}$  significantly reduces the number density of the most massive black holes at higher redshifts ($z > 15$).   An interesting (although weak) signature  of the presence of ${\bf \vbc}$ may be the modulation at BAO scale  of the observed high-redshift quasars, since velocities may reduce masses and thus luminosity of the supermassive black holes \cite{Tanaka:2013}.

When the second scenario is considered, ${\bf \vbc}$ can facilitate the formation of supermassive black holes via direct collapse \cite{Tanaka:2013b, Latif:2013} by letting the halos  evolve to higher masses before the pristine gas is accreted. In rare patches with $\vbc>2\sigma_{bc}$, the most massive halos at $z \sim 30$ could reach the critical temperature for direct collapse before forming stars by atomic cooling, assuming central densities are high enough to collisionally dissociate $H_2$. However, it was pointed out recently \cite{Visbal:2014} that creating such a heavy halo (with $T \sim 10^4$ K) before cooling becomes efficient is not a sufficient condition for forming a black hole via direct collapse. It was shown \cite{Visbal:2014} that although H$_2$ formation may be delayed, molecular hydrogen will eventually form at high enough densities in the center of the halo leading to efficient cooling, fragmentation and thus star formation. It seems that the only obvious way to avoid molecular cooling (in the absence of Lyman-Werner radiation), is for gas to reach even higher density to cause collisional dissociation of molecular hydrogen before cooling occurs. However, it appears  that the minimum core entropy, set by the entropy of the intergalactic medium when it decouples from the CMB, prevents this from occurring for realistic halo masses.


\section{Signature  of the Velocities in Detectable Cosmological Signals}
\label{Sec:observ}

Precision cosmological measurements today allow us to learn about our cosmic ``neighborhood'' from observations of galaxies, quasars, gamma ray bursts, supernovae etc. at redshifts bellow $\lesssim 10$ as well as about the very young Universe at the moment of decoupling of the CMB at $z\sim 1100$. In the upcoming future,  we will most likely be able to complete the picture by performing a three-dimensional mapping of the signal from neutral hydrogen  (the most abundant element in the Universe after hydrogen atoms formed at $z\sim 1100$ and before most of it was re-ionized at $z\sim 10$). Measuring photons with wavelength which corresponds to the redshifted  21-cm transition of neutral hydrogen, one could in principle  learn about astrophysics and cosmology at the redshifts $10\lesssim  z \lesssim 1100$.

Interestingly enough, all the above-mentioned types of measurements may be affected by ${\bf \vbc}$, as we discuss in this section, but the most interesting potential  signature of the relative velocities is expected to be found in the redshifted 21-cm  signal from the epoch when first stars  form $20\lesssim z\lesssim 50$ \cite{ OLeary:2012, Visbal:2012, Fialkov:2013, Fialkov:2014, AliHaimoud:2013}, as this is exactly the range of redshifts where the impact of ${\bf \vbc}$ on structure formation and radiative backgrounds is strongest. At that epoch, in fact, the signal produced by neutral hydrogen  in the intergalactic medium couples to the radiation emitted by first galaxies; thus if the first galaxies resided in light halos of M$\sim 10^5-10^6$ M$_\odot$, the radiative backgrounds emitted by the first stars would bear the signature of ${\bf \vbc}$, which then would  be imprinted in the 21-cm signal. Other possible signatures of $\vbc$ include effects on the B-mode polarization of the CMB \cite{Ferraro:2012} (although expected to be non-significant),  reionization \cite{Bittner:2011, Pritchard:2012},  the location of BAO peaks measured in large scale structure surveys \cite{Yoo:2013, Dalal:2010}, the abundance of low-mass satellite galaxies \cite{Bovy:2013}, quasars \cite{Tanaka:2013, Tanaka:2013b} (which we have already discussed in the previous section), etc. In this section we mention all the observable effects of ${\bf \vbc}$ that were discussed in the literature so far.

\subsection{Redshifted 21-cm signal of neutral hydrogen}

We start with discussing the imprints of ${\bf \vbc}$ in the 21-cm signal, which is a potentially powerful probe of the early Universe. Because the 21-cm signal is sensitive to many astrophysical and cosmological parameters, making reliable predictions for it is very challenging. Despite this, we can show that the modulation in the signal by  ${\bf \vbc}$, which we discuss in this section, is expected to manifest itself on large scales, tracing the BAO fluctuations. 

The cosmological 21-cm signal is observed with the CMB as an all-sky bright source. Emission and absorption of the $\lambda_{21} = 21$ cm wavelength by hydrogen atoms located at redshift $z_H$ from the background radiation deforms the initial Black Body spectrum of the CMB at the wavelength $\lambda = \lambda_{21}(1 + z_H)$ cm.  Photons with  wavelength $\lambda_{21}$ are emitted by hydrogen atoms if the effective temperature of the spin-flip transition (called the spin temperature, $T_S$) is hotter than the CMB, which leads to an increment in the measured spectrum at the wavelength $\lambda$ with respect to the initial Black Body. On the other hand, if the spin temperature is colder than that of the CMB, hydrogen atoms absorb $\lambda_{21}$ photons from the CMB, which leads to a trough in the observed spectrum at the corresponding wavelength. According to the present understanding, the main features of the  expected global signal are: (1) the redshifted 21-cm signal from very early epochs ($z > 200$) is negligible  due to thermal coupling of the gas to the CMB\footnote{However, an enhancement is expected in the global intensity and fluctuations of the 21-cm  signal around the epoch of hydrogen recombination. Such an enhanced signal could be produced due to  coupling of the 21-cm line of neutral hydrogen to the Ly-$\alpha$ background generated by recombining hydrogen atoms \cite{Fialkov:2013b}.}; (2) after the gas thermally decouples from the CMB at $z\sim 200$, it cools adiabatically due to the cosmic expansion at a rate which is faster than the rate at which CMB cools.  The spin temperature of the 21-cm line at this epoch is driven to the temperature of the gas, $T_K$, by  hydrogen-hydrogen collisions. Since the gas at that epoch is  colder than the radiation, the observed signal  would be seen in absorption; (3) when the gas rarefies enough to make the collisional coupling inefficient, the spin temperature is driven to that of the CMB, and the global signal vanishes; (4) after the first stars start to shine, the starlight couples the spin temperature to the color temperature of the Ly-$\alpha$ radiative background \cite{Madau:1997}, $T_c$ (which during most epochs closely follows the gas temperature),  via the Wouthuysen-Field (WF) effect \cite{Wouthuysen, Field}, i.e.,  absorption and re-emission of Ly-$\alpha$ photons. At first, the gas is colder than the CMB and the signal is seen in absorption; however, slowly X-ray photons emitted mainly by stellar remnants (though the exact heating mechanism at high redshifts is yet unconstrained by observations) heat the gas above the temperature of the CMB, and the 21-cm signal is seen in emission. 

The observable  brightness temperature  of the cosmological 21-cm signal relatively to the CMB is parametrized as follows \cite{Pritchard:2012}
\begin{equation}
\label{eq:T21}
\delta T_b \approx 27 x_{HI}(1+\delta)\left(\frac{\Omega_bh^2}{0.023}\right)^{-1/2}\left(\frac{0.15}{\Omega_mh^2}\frac{1+z}{10}\right)^{-1/2}\left(1-\frac{T_{CMB}}{T_S}\right)\left[\frac{H}{H+\partial_rv_r}\right],
\end{equation}
in mK units. Here $x_{HI}$ is the neutral fraction of hydrogen,  $\Omega_m$ is the fractional overdensity in the total amount of matter (including baryons and dark matter),  and $\partial_rv_r$ in the final term is the velocity gradient along the line of sight which gives rise to redshift space distortions.  The spin temperature  reads 
\begin{equation}
T_S^{-1} = \frac{T_{CMB}^{-1} +x_c T_K^{-1} +x_\alpha T_c^{-1}}{1+x_c+x_\alpha}
\end{equation}
where $x_c$ and $x_\alpha$ are the collisional coupling and the WF coupling, respectively.  The spin temperature, in general, is a  complicated function of many parameters and depends on the temperatures of the radiation and gas and on the  intensities of the Ly-$\alpha$,  X-ray, and LW radiative  backgrounds, which are created by the non-local population of stars and fluctuate from place to place \cite{Barkana:2005, Pritchard:2007}. Therefore, the 21-cm background is also  inhomogeneous, and inherits spatial fluctuations from each one of its components. In addition to that, fluctuations in the 21-cm signal are seeded by fluctuations in  initial conditions for structure formation, including  large scale overdensities,  relative velocities, and fluctuations in the neutral fraction. For more details on the 21-cm signal see, for instance, a review by  J. R. Pritchard and A. Loeb (2012) \cite{Pritchard:2012}.

\subsubsection{Signal from the era of primordial star formation}

Numerical simulations, as well as semi-analytical calculations, face a great difficulty when trying to simulate the population of first stars and thus the 21-cm signal emitted at the epoch of primordial star formation $10\lesssim z \lesssim 50$. On one hand, it is hard to make reliable predictions using only semi-analytical methods, since the neutral hydrogen signal strongly depends on non-local radiative backgrounds, and since we need to account for highly non-linear processes such as star formation, feedbacks etc.  On the other hand, numerical simulations, in which the non-linear physics can be treated properly, have to face the  difficulties resulting from having to  resolve the then-typical tiny galaxies while at the same time capturing the global distribution of rare objects. Relative velocities further complicate this picture adding large-scale bias to star formation. In other words, cosmological simulations that cover the complete range of scales are not currently feasible. To overcome this difficulty and properly simulate the 21-cm signal on large scales it is convenient to use hybrid computational method \cite{ Fialkov:2012, Visbal:2012, Fialkov:2013, Fialkov:2014}. An alternative approach was taken in a papers by N. Dalal, U.-L. Pen and U. Seljak (2010) \cite{Dalal:2010} and  by M. McQuinn and R. M. O'Leary (2012) \cite{McQuinn:2012}  in which semi-analytical estimates were used, where the later calculation is calibrated to dedicated  numerical simulations \cite{OLeary:2012}.

The authors of the first study, in which the effect of ${\bf \vbc}$ on the  21-cm signal at the epoch of primordial star formation ($z\sim 20$) was computed, were N. Dalal, U.-L. Pen and U. Seljak (2010) \cite{Dalal:2010}. They showed for the first time using  semi-analytical methods that ${\bf \vbc}$ imprints the enhanced BAO signal in the neutral hydrogen signal.  However in this first pioneering paper several simplifying assumptions, which are now relaxed in more advanced studies \cite{Fialkov:2013, Fialkov:2014,  Fialkov:2014c}, were made. For instance, the authors accounted for fluctuations in the 21-cm signal sourced by  Ly-$\alpha$ radiative background only, ignoring X-rays and LW. In addition, out of the effects of ${\bf \vbc}$ discussed in section \ref{Sec:structure} the suppression in the gas content of halos was the only one included, while the suppression of halo abundance as well as the increment of $M_{cool}$ were ignored; moreover, a simplified prescription for the filtering mass was used. As a result, the overall effect of ${\bf \vbc}$ on the 21-cm signal was underestimated in this study.  

To ameliorate the predictions for the effect of relative velocities, in the paper by E. Visbal et al. (2012) \cite{Visbal:2012} hybrid numerical methods were used to calculate the large scale  fluctuations in the 21-cm signal seeded by inhomogeneous gas temperature  at redshift $z = 20$ (assumed to be the redshift of heating transition, i.e. the redshift at which $T_K = T_{CMB}$ in average over the universe), and biased by the relative velocities and large-scale overdensities. In this work, all three effects of ${\bf \vbc}$ on structure formation (suppression of halo abundance, suppression of gas fraction and boost in the cooling mass) were accounted for; the effect of  fluctuations in Ly-$\alpha$ background  was assumed to be saturated by redshift $z = 20$, meaning that due to the strong intensity of the Ly-$\alpha$ photons ($x_\alpha>>1$) the spin temperature was tightly coupled to the gas temperature $T_S\approx T_c\approx T_K$.  These assumptions led to a simplified version of eq. \ref{eq:T21}  and allowed to efficiently explore the effect of  ${\bf \vbc}$ on  the fluctuations in the 21-cm signal seeded primarily by inhomogeneous heating. These fluctuations appear to be enhanced due to the fact that now the signal probes more rare halos (either atomic-cooling halos or molecular cooling biased by ${\bf \vbc}$).  The only additional source of fluctuations in this case are the  density fluctuations.  In this work two cooling criteria were considered in order to gain understanding of how much the LW negative feedback could affect the imprints of ${\bf \vbc}$: one including molecular hydrogen cooling halos, and a second which excluded this mode.  Clearly, the expectation was to find a clear  signature of ${\bf \vbc}$ in the former scenario which includes $\gtrsim 10^5$ M$_\odot$ halos unaffected by LW feedback. 
The  dimensionless power spectrum, as reported in E. Visbal et al. (2012) \cite{Visbal:2012} for the range of wavenumbers k$\sim 0.03 - 0.3$ Mpc$^{-1}$, appears to be overall flat with  strong imprints of BAO seeded by ${\bf \vbc}$ in the scenario with  molecular cooling. On the contrary, the second history without molecular cooing halos does not show any visible traces of BAO. This leads to an interesting observation: the detection of strong BAO features in the 21-cm power spectrum may constrain our models of high-redshift cooling mechanisms.

\begin{figure}[t]
{\center
\includegraphics[width=4in]{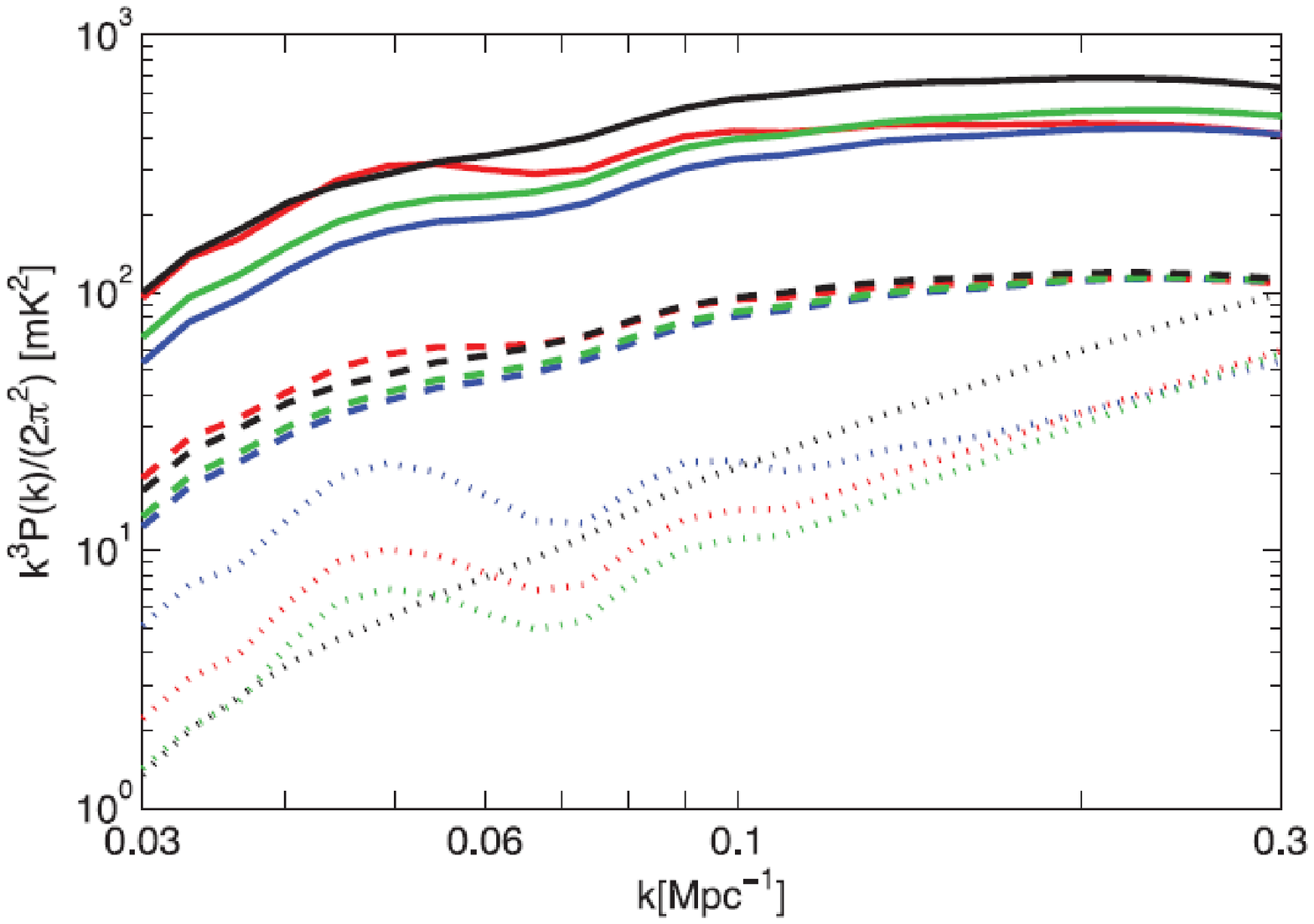}
\includegraphics[width=5.0in]{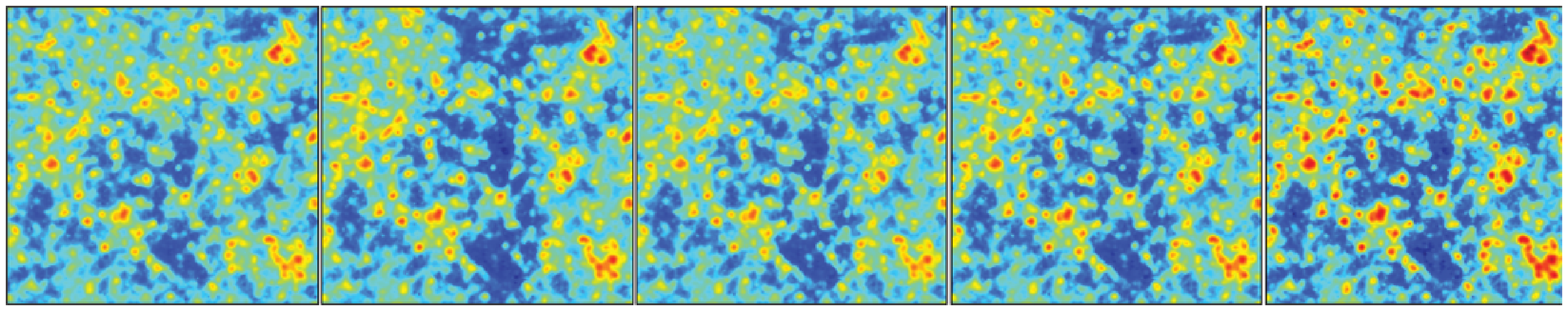}
\caption{\label{fig:21LW} Top: Power spectra of the 21-cm brightness temperature for no feedback (red), weak feedback (blue), strong feedback (green) and saturated feedback (black)  including ${\bf \vbc}$, at redshifts $z = z_0$ (dashed), $z = z_0 + 3$ (solid) and $z = z_0 + 9$ (dotted). Bottom: The 21-cm brightness temperature $T_b$ (relative to the cosmic mean in each case) in mK, shown at $z_0 + 3$, where $z_0$ is the redshift of heating transition. The slices of the 21-cm signal correspond to the inital conditions for density and relative velocity shown on figure \ref{fig:IC}. The cases shown (from left to right) refer  to: no feedback, no ${\bf \vbc}$; no feedback; weak feedback; strong feedback; and saturated feedback (where the last four include the streaming velocities). The cosmic mean values (which have been subtracted from the maps) are: $\bar{T_b}$	 = -81, -89, -73, -88 and -108 mK, respectively. The colors correspond to $T_b$ from -100 mK (blue) to 150 mK (red). The figure is adopted from  A. Fialkov et al. (2013) \cite{Fialkov:2013}. }}
\end{figure}

In reality, the dependence of the 21-cm signal on ${\bf \vbc}$ is most likely much more complex then described in E. Visbal et al. (2012) \cite{Visbal:2012}, due to the coupling of the 21-cm signal to non-local and inhomogeneous radiative backgrounds, unaccounted for in the aforementioned work,   which are biased by the supersonic relative velocities. The effect of ${\bf \vbc}$ together with the inhomogeneous time-evolving LW feedback and X-ray background on the 21-cm signal was studied in A. Fialkov et al. (2013) \cite{Fialkov:2013}. Examples of slices of the 21-cm signal affected by ${\bf \vbc}$ with different scenarios for LW feedback are shown in figure \ref{fig:21LW}. The slices of a simulation box, shown in this figure, correspond to the set of initial conditions for large-scale overdensity and $\vbc$  shown earlier in fig. \ref{fig:IC}. As a snapshot of a universe for each model would look very distinctive for a fixed redshift mainly due to the difference in the heating history (delayed by both ${\bf \vbc}$ and LW), the authors chose to show the plots at a redshift related to heating transition of the simulated universe (i. e., the moment in cosmic history when the average gas temperature equals to the CMB temperature $<T_K> = T_{CMB}$),  which happens respectively at $z_0 =$  16.6, 15.2, 14.6, 14.2 for the cases of no LW feedback, weak feedback, strong feedback and saturated feedback (all cases with ${\bf \vbc}$), while in the case when  both ${\bf \vbc}$ and feedback are ignored $z_0 $ was found to be 17.4.  To be precise, the snapshots are shown at the redshift $z_0+3$, since as was seen in the simulations \cite{Fialkov:2013},  this is roughly the redshift at which heating fluctuations in the 21-cm signal are maximal assuming a soft power-law heating spectrum\footnote{Note that the character of the spectral energy density  of heating fluctuations can have a strong impact on the expected 21-cm signal \cite{Fialkov:2014b, Fialkov:2014c}.}. With the overall normalization factored out, the physical differences become clearer, in particular, one can still see a clear signature of ${\bf \vbc}$ in the maps corresponding to the two realistic feedback cases,  namely the weak and strong ones. However, as anticipated, in these cases the signature of ${\bf \vbc}$ is not as  strong as in the  case with no feedback, i.e., in the case when there is no suppression in the abundance of molecular hydrogen. To understand the comparison between the weak and the strong feedbacks, we remind that the velocities cause a very strong suppression of star formation up to a halo mass M$\sim 10^6$     M$_{\odot}$, but above this critical mass the suppression and its dependence on halo mass weaken considerably. This and other features can be seen more clearly and quantitatively in the power spectra of the 21-cm signal \cite{Fialkov:2013} shown in  figure \ref{fig:21LW}. Initially, the 21-cm fluctuations from inhomogeneous heating rise with time as the gas heats up, and this happens first in  regions with a high stellar density. Then, at the more advanced stages of heating, the 21-cm fluctuations faint away, since the 21-cm intensity becomes independent of the gas temperature once the gas is much hotter than the CMB, i.e., the heating fluctuations saturate at low redshifts\footnote{Note that in reality the redshifted 21-cm  signal is most likely even more complicated due to the fluctuations in the ionization fraction of neutral hydrogen, $x_i$ \cite{Fialkov:2014b}.}.

\subsubsection{Neutral hydrogen signal from the dark ages}

Observations of the redshifted 21-cm signal from the dark ages $z\gtrsim50$ (before star formation has started) are much more challenging than from lower redshifts. The  main reason is that the signal at wavelengthes of $\lambda_{21}(1+z)\gtrsim 10 $ meters are reflected by the ionosphere of the Earth, which means that their detection from the ground is impossible. A space or lunar array would therefore be required to observe the signal from the dark ages \cite{Weiler:1988, Wolt:2012, Lazio:2011}. Fluctuations in the redshifted 21-cm signal from the dark ages   mainly arise from fluctuations in baryon density $\delta_b$ \cite{Loeb:2004, Lewis:2007}, which were  still linear on most  scales at that epoch. Measuring them at such high redshifts may thus  provide in the future  a high-precision cosmological probe, which would allow to examine  the pristine fluctuations in energy  density contrast as seeded by cosmological inflation or other alternative mechanism.

\begin{figure}[t]
{\center
\includegraphics[width=3.4in]{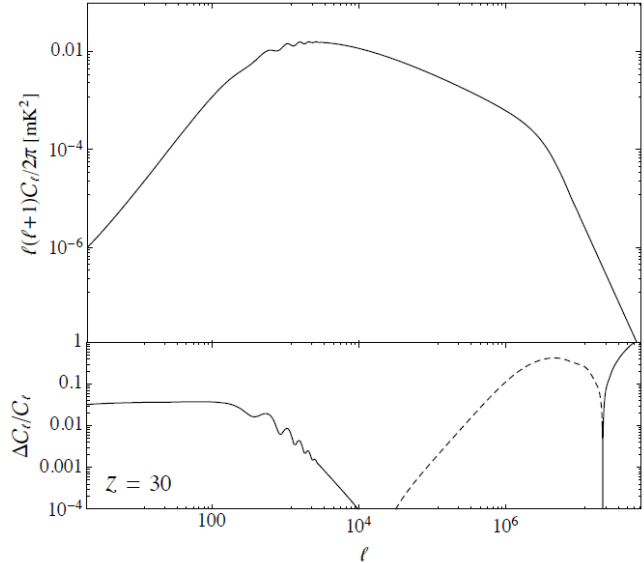}
\caption{\label{fig:Dark} Top panel: 21-cm angular power spectrum at redshift 30 without relative velocity corrections. The relative correction to the angular power spectrum  when accounting for the relative velocity effect are shown in the bottom panel: solid lines represent an enhancement and dashed lines a suppression relatively to the spectrum shown on the top panel. The figure is adopted from  Y. Ali-Haimoud, P. D. Meerburg  and S. Yuan (2013) \cite{AliHaimoud:2013}.}}
\end{figure}

The effect of the relative velocity on the 21-cm brightness temperature fluctuations from the dark ages, i.e., excluding star formation,  was recently computed by Y. Ali-Haimoud, P. D. Meerburg  and S. Yuan (2013), and reported in a recent paper \cite{AliHaimoud:2013} where the authors solved a set of modified fluid equations similar to eqs. \ref{eq:sys2} and accounted for fluctuations in gas temperature and free electron fraction sourced by the density fluctuations. The main finding of this study is that the non-linear effects due to ${\bf \vbc}$ mildly affect the power of the 21-cm fluctuations  at a large range of scales ($0.001\lesssim k \lesssim 2000$ Mpc$^{-1}$) and at redshifts $20\lesssim z\lesssim 200$ with the maximal effect being  at $z\sim 30$. Previous expectations   for the two-dimensional angular power spectrum $C_l$ of the 21-cm signal \cite{Loeb:2004, Lewis:2007} are  shown in the top panel of figure \ref{fig:Dark}, while the bottom panel of the figure shows the relative modification     of the angular power spectrum due to ${\bf \vbc}$ at $z= 30$ for a wide range of scales. The signal is enhanced by up to tens percent at large scales  $0.005\lesssim k \lesssim 1$ Mpc$^{-1}$ (corresponding to  angular scales of $l\lesssim 10^4$ at $z = 30$) due to the non-linear dependence of the 21-cm brightness temperature on the underlying gas density and temperature; the signal is also mildly amplified  at very small scales of $k \gtrsim 2000$ Mpc$^{-1}$ (angular scales of $l\gtrsim 5\times 10^7$ at $z = 30$)  by resonant amplification of acoustic waves. At intermediate scales the signal is instead suppressed by up to tens percent. In principle, the results of this paper show that, at least to the order considered by the authors, some of the small-scale power leaks to larger scales, suggesting that  extracting cosmological data from the 21-cm signal may be slightly easier than expected.

One should keep in mind, however, that such a small enhancement/suppression of the power spectrum is unlikely to be detected in the near future as neither present nor next generation 21-cm experiments will have the sensitivity to detect it. In addition, as was recognized by the authors, the signal is expected to be maximal at $z\sim 30$, at which point the first stars had most likely already formed \cite{Fialkov:2012, Naoz:2006} and ``contaminated'' the cosmological signal.

\subsection{Effects of ${\bf \vbc}$ during reionization}
\label{Sec:reion}
Cosmological reionization is believed to be one of the major phase transitions through which the observable Universe underwent during its evolution.  Once stars formed in the pristine   neutral gas, they  emitted ultraviolet radiation which in turn ionized the surrounding gas. Astrophysical evidences for this phase transition include the measurements of the total Thompson scattering optical depth due to reionization from CMB anisotropies \cite{Ade:2013}, the specific luminosity density of the observed galaxy population at $6<z<8$ \cite{Grazian:2012, Finkelstein:2012}, and the Gunn-Peterson \cite{Gunn:1965} absorption troughs in the spectrum of high redshift bright sources such as quasars, Lyman-$\alpha$ emitters and $\gamma$-ray bursts \cite{Fan:2006, Mortlock:2011, Willott:2007, Nakamura:2011, Robertson:2012}. Starting when the very first sources of  ionizing photons appeared, the epoch of reionization was concluded around $6\lesssim z \lesssim 12$, according to the latest observational tests \cite{Mortlock:2011, Ade:2013,  Grazian:2012, Finkelstein:2012, Nakamura:2011, Robertson:2012}. In addition, duration of reionization has been constrained to $\Delta z<7.9$ by measuring the kinetic Sunyaev-Zeldovich effect \cite{Zahn:2012}. More precise observations which involve actual measurements of the ionized fraction of hydrogen by mapping the intensity of its redshifted 21-cm line are  currently on the way, with experiments such as LOFAR and MWA presently taking data.

Due to the lack of direct observations at present, our understanding of the process of reionization and the sources that contributed most intensively to the ionizing background is still limited.  The cosmological background of ionizing radiation is dominated today by emission from quasars,  however, at higher redshifts the observed abundance of bright quasars declines sharply \cite{Fan:2001, Srbinovsky:2007, Hopkins:2007}. As a result, the contribution of quasars lessens  at higher redshifts  at the point that  by redshift $z\sim 3$ quasars and stars contribute equally to the ionizing background, and by redshift $z\sim 6$ the contribution of quasars becomes sub-dominant \cite{Volonteri:2009}. However, none of the existing known population of sources is able to account alone for the intensity of the ionizing background needed to ionize cosmic hydrogen by $z\sim 6$ \cite{ Grazian:2012, Finkelstein:2012, Fontanot:2014}. Thus, in many theoretical scenarios, star formation in  faint galaxies beyond the sensitivity  of present day observatories is introduced in order to produce a sufficiently ionizing background.

Although an answer is  not yet known with certainty, a recent work \cite{Munoz:2011}  suggested that the reionization sources  were dominated by early galaxies of virial temperatures above the hydrogen cooling threshold of $10^4$ K, corresponding to masses $\gtrsim 10^8$ M$\odot$.  As was discussed above, this range of masses is not sensitive to $\vbc$ and therefore the redshift of reionization, distribution of ionized patches, and characteristic size of ionized bubbles are not expected to be strongly correlated with ${\bf \vbc}$ in this scenario. On the other hand, if molecular cooling does turn out to be dominant for sources that produce most of the ionizing photons, the ${\bf \vbc}$ effect would delay reionization in an inhomogeneous manner, with average delay estimated to be $\Delta z \sim 2$ \cite{Bittner:2011}. This delay would be degenerate with uncertainties in the star-forming efficiency f$_\star$, the escape fraction $f_{esc}$ of ionizing photons from galaxies, and other astrophysical parameters such as the initial mass function of stars which, however, are not expected to have the same stochasticity as the delay introduced by  ${\bf \vbc}$. 

Another possibility is that light halos of M$\sim 10^6$ M$_\odot$, which are biased by ${\bf \vbc}$, may have pre-ionized the intergalactic medium, imprinting the effect of the relative velocities into the fraction of ionized gas  at higher redshifts before most of the reionization occurs.  In this case, the large-scale modulation of ${\bf \vbc}$ would lead to  weak  large scale fluctuations in the total optical depth to reionization \cite{Su:2011}.

Interestingly, the anisotropy in the Thompson scattering optical depth, or equivalently in the  free electron fraction, would in principle lead to a rise of an unexpected contribution to the B-mode polarization of the CMB \cite{Ferraro:2012}, in addition to the contributions from gravitational lensing by large scale structure \cite{Zaldarriaga:1998} and those due to the primordial tensor fluctuations \cite{Polnarev:1985}. The two-dimensional B-mode angular power spectrum $C_l$ computed by S. Ferraro, K. M. Smith and C. Dvorkin (2012) \cite{Ferraro:2012} has a characteristic shape with acoustic peaks at multipoles $l = 200$, $400$, etc. The amplitude of this signal is a free parameter which is related to the dependence of the ionization fraction on ${\bf \vbc}$ during the epoch of reionization, the scenario for reionization and other model parameters. Assuming   reionization models in which hydrogen is reionized promptly at $z \sim 10$  (e.g., instantaneous reionization), the amplitude of the B-mode signal from ${\bf \vbc}$ is undetectably small. In particular it is three-four orders of magnitude below the B-mode signal reported by BICEPII Collaboration \cite{BICEP:2014}.  

Another effect related to ${\bf \vbc}$ that can affect reionization and that has been explored in literature is its impact on light halos that do not form stars. In standard cosmology, i.e., not taking into account the possibility of ionization by dark matter annihilation and other exotic scenarios, such halos do not host ionizing sources, and thus the gas inside of them remains largely unionized. Neutral gas can block ionizing radiation, and induce an overall delay in the expected progress of reionization \cite{Shapiro:1987, Shapiro:2004,  Iliev:2003, Iliev:2005, Barkana:2002, McQuinn:2007}. In scenarios involving a non-vanishing ${\bf \vbc}$, the amount of gas in light star-less halos is significantly suppressed (as discussed in section \ref{Sec:structure}), and therefore  such halos do not block the ionizing radiation as efficiently as they do in the case with ${\bf \vbc} = 0$ \cite{Naoz:2012}. Naturally, the efficiency of this phenomenon is expected to be modulated by the local magnitude of $\vbc$.

\subsection{Imprints of the velocities in present day galaxies}
Present day star formation mainly happens in massive dark matter halos of $\sim 10^{10}-10^{14}$ M$_{\odot}$ as estimated, for instance, combining halo abundance matching and  observations of stellar masses \cite{Papastergis:2012}, which are not expected to be strongly affected by ${\bf \vbc}$.  In addition, $\vbc$ decays with redshift, meaning that the mean large-scale relative velocity today is  only $\sim 0.03$ km s$^{-1}$, which makes the detection of its effect  in the present day galaxies even more difficult. Therefore the direct impact of ${\bf \vbc}$ on star formation after reionization ($z\lesssim 10$) is expected to be negligible and naively we would not expect to detect any ${\bf \vbc}$-related signature today. However, as is argued in the literature, and as we review in this section, the relative motion can still impact the low-redshift structure formation indirectly. The signatures of  $\sim 10^6$ M$_\odot$ halos could persist even in late-time observables; for instance, if reionization is patchy on $\sim 100$ Mpc scales due to the effects of ${\bf \vbc}$ at higher redshifts  mentioned in section \ref{Sec:reion}, the subsequent star formation history inside patches that reionize earlier should lag from patches that reionize later. This could lead to spatial variations  on scales of order the BAO scale in star formation rates, metallicity and other factors at lower redshifts.  Other possible low-redshift signatures  of ${\bf \vbc}$ are its possible impact on the formation of globular clusters (which we mentioned in section \ref{Subsec:Halos}) and its effect on satellite galaxies and on the location of BAO peaks measured from low-redshift massive galaxies,  which we discuss in the following.
 
\subsubsection{Resolving the missing satellite problem}

One of the examples of low-redshift astrophysical objects that can keep traces of ${\bf \vbc}$ is given by luminous low-mass satellite galaxies around larger galaxies such as the Milky Way and Andromeda. The abundance of the low-mass bright galaxies is known to be in tension with the predictions of the canonical $\Lambda$CDM cosmology which assumes  hierarchical structure formation \cite{Weinberg:2013}. More precisely there are two problems: the inconsistency of the observed core of dark matter halos with the predicted cuspy profile, and the too little number of observed satellite  galaxies compared to the amount  predicted by ``dark matter only" simulations \cite{Kauffmann:1993, Klypin:1999, Moore:1999}. The latter problem, usually referred to as  the missing satellite problem, can be (partially) solved by modifying the theory, e.g., invoking   warm dark matter scenarios \cite{Polisensky:2011}, or modifications to the cosmological inflation paradigm in which  small-scale power is suppressed with respect to the cold dark matter scenario \cite{Kamionkowski:2000}. However, most likely the missing satellite problem can be solved by correctly treating the physics of baryons in  cosmological simulations of structure formation, most of which still ignore baryons out of simplicity. For instance, sub-halos could be empty of gas and  stars, in which  case it would be much harder to detect them using galaxy surveys. Hinting to the fact that
this may be the right explanation, a dedicated search using Sloan Digital Sky Survey found $\sim 15$ new  satellites with very low luminosities \cite{Willman:2005, Belokurov:2007}.  Moreover, to explain why the low-mass sub-halos would be devoid of stars several arguments can be invoked. For instance,   gas accretion into small galaxies is expected to be suppressed due to photo-heating by ultraviolet ionizing background, which should decrease star formation \cite{Bullock:2000, Benson:2002, Somerville:2002}. In this case, star formation in small galaxies of total mass M$\lesssim 10^8$  M$_\odot$ becomes very inefficient after reionization \cite{Madau:2008}. Additionally, the gas can be stripped away from the shallow potential wells of the satellites by   supernovae winds. However, to fully explain current observations within $\Lambda$CDM, additional suppression of star formation before reionization is needed \cite{Koposov:2009}. Finally, the baryon-free dark matter satellites could be formed thanks to the effects of ${\bf \vbc}$ \cite{Naoz:2014}, which we mentioned in section \ref{Subsec:Halos}.

\begin{figure}[t]
{\center
\includegraphics[width=3.4in]{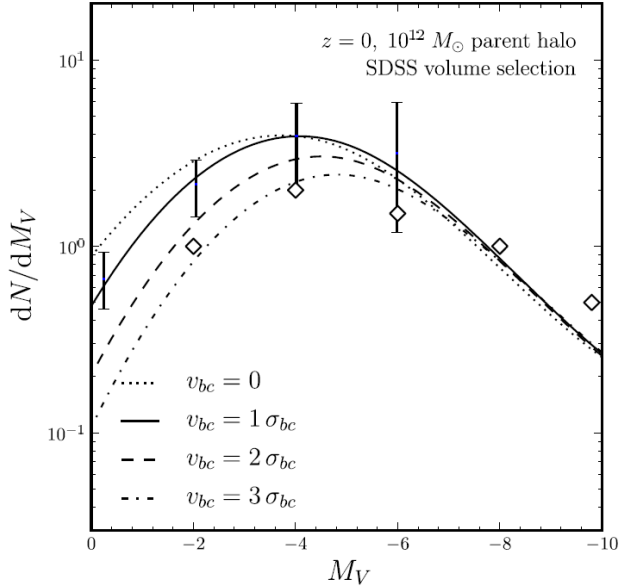}
\caption{\label{fig:lum} Luminosity function of satellite galaxies of the Milky Way observed by the SDSS. The diamond data points are taken from Koposov et al. (2009) \cite{Koposov:2009}. The error bars on the $\vbc = 1\sigma_{bc}$ model curve show the $68\%$ percent spread in the merger trees at luminosities where the predictions with different $\vbc$ differ; this spread is similar for all four model curves. The error bars are highly correlated; for example, the correlation between the MV = −4 and MV = −2 error bars is 0.75. For this figure star formation efficiency of  $f_\star = 0.01$ was assumed. The figure is adopted from  J. Bovy and C. Dvorkin (2013) \cite{Bovy:2013}.}}
\end{figure}

Recently, an additional mechanism to suppress star formation  in satellite galaxies was proposed  by  J. Bovy and  C. Dvorkin (2013) \cite{Bovy:2013} who use the relative motion between baryons and dark matter wells to suppress the number density of halos with masses in the range $10^6<$ M $< 10^8$ M$_\odot$  that host luminous galaxies today. This is potentially important, as the earlier unaccounted effect of ${\bf \vbc}$ may help to resolves most of the tension between the observed and predicted number of low-mass satellites in the Milky Way. The satellite halos must form a large number of their stars before reionization because the photo-heating effect shuts down star formation at lower redshifts. The authors apply  merger-tree simulations using the Extended Press-Schechter formalism  for a parent halo of mass $M = 10^{12}$ M$_\odot$ to estimate the satellite luminosity function at $z=0$. In their work, the gain in   stellar mass after reionization is calculated taking into account the suppression of star formation by photoionization. The authors show that  ${\bf \vbc}$ can suppress the stellar mass function by typically $\sim 50\%$ at the low-luminosity end (the effect of ${\bf \vbc}$ on the luminosity function is shown in figure \ref{fig:lum}). The origin of the main contribution to the suppression appears to be the decreased  gas accretion onto these halos, rather than the suppressed  halo abundances (${\bf \vbc}$ typically suppresses the number of dark matter halos by 10 to 20 percent in the range $10^5 < M < 10^7$ M$_\odot$ at z = 11). The main conclusion of this paper is that in the case of  $\vbc = 1 \sigma_{bc}$, there is almost no discrepancy between  their predictions and the number counts of the SDSS\footnote{http:$//$www.sdss.org$/$} survey at the faint end, i.e., at luminosity of M$_V>-6$. Note though, that these authors do not account neither for the boost in the minimal cooling mass eq. \ref{Mcool0} nor for the fatal effect of Lyman-Werner photons on star formation. Including these effects may render the results to be less impressive.

\subsubsection{Signature of ${\bf \vbc}$ in massive low-redshift galaxies}
\label{Sec:LRG}

It is believed that the modulation of early halos by  ${\bf\vbc}$ affects the formation of higher-mass low-redshift galaxies, through, for instance, inhomogeneous metal abundance or supernovae rate \cite{Yoo:2011}. If unaccounted for,  the effect of the relative velocities  can bias the measurements of the BAO peak positions, affecting the estimate of the dark energy equation-of-state \cite{Yoo:2013}. The  impact of ${\bf \vbc}$ on  spatial clustering of low-redshift massive galaxies was studied by J. Yoo, N. Dalal and U Seljak (2011) \cite{Yoo:2011}, J. Yoo and U. Seljak (2013) \cite{Yoo:2013}, and Z. Slepian and D. Eisenstein (2014) \cite{Slepian:2014}. Since the connection between the feedback and late-time galaxies distribution is not known in detail, in the presence of ${\bf \vbc}$, the large-scale fluctuations of collapsed objects (galaxies) can be modeled with two non-linear galaxy biases, $b_1$ and $b_2$ and the relative velocity bias parameter $b_r$
\begin{equation}
\delta_g(x) = b_1\delta_m(x)+\frac{b_2}{2}\left[\delta_m^2(x)-\sigma_m^2\right]+b_r\left[\vbc^2(x)-\sigma_{bc}^2\right],
\end{equation}   
 where $b_r$ characterizes our uncertainty in the  amplitude of the remaining relative velocity effect in low-redshift galaxy populations, and the nonlinear galaxy bias parameters ($b_1$, $b_2$) are the lowest order coefficients of the local matter density expansion that relate to the galaxy number density. This lowest-order parametrization allows to study the impact of ${\bf \vbc}$ on the large scale clustering properties of galaxies. For instance, for a fiducial choice of the bias parameters ($b_1 =1$, $b_1 =2$ and $b_r =0.04$)  the effect of ${\bf \vbc}$ on the galaxy power spectrum $<\delta^*_g({\bf k})\delta_g({\bf k})>$ appears to be more than $10\%$  on large scales of k $\lesssim  0.01$ h Mpc$^{−1}$ (throughout this subsection we are using $h = 0.7$), and declines as k$^{-4}$ on smaller scales at which BAO are usually measured \cite{Yoo:2011}. In addition, the coupling to ${\bf \vbc}$ generates a large scale bispectrum of galaxies, which peaks around the scale k $\sim 0.03$  h Mpc$^{-1}$.  More importantly, the relative velocities may introduce a shift  in the position of BAO peaks  measured from the galaxy distribution, which can reach few percents if the bias parameter is large ($b_r\sim 0.2$) \cite{Yoo:2011}.  A shift  of the position and a broadening of BAO peaks also occurs due to various non-linear effects beside the one  discussed here related to ${\bf \vbc}$. Such non-linear effects can be accounted for and marginalized over in measuring the positions of BAO peaks by applying a template power spectrum \cite{Seo:2008}, which has to be consistently  modified to include the effects of ${\bf \vbc}$ as well \cite{Yoo:2013}.

Given that amplitude of the effect  ${\bf \vbc}$ may have on the clustering of massive galaxies today is expected to be very small, of the order of $\lesssim 1\%$, a direct measurement of it would be difficult.  However if there is another available sample of galaxies that does not trace ${\bf \vbc}$, e.g., present day young star forming low-mass galaxies, correlating the two samples would help to single out the effects of ${\bf \vbc}$. The fact that the two galaxy samples trace the same matter distribution allows to construct a combination of the two data sets in which the contribution of matter fluctuations disappears \cite{Yoo:2013}.  In the  power spectrum obtained from such a data set, the effect of ${\bf \vbc}$ is expected to be the dominant feature  at large scales k $\lesssim 0.1$ Mpc$^{-1}$. This procedure was followed by J. Yoo and U. Seljak (2013) \cite{Yoo:2013}  who applied it to real galaxies using the publicly available spectrum measurements \cite{Anderson:2012} of the CMASS galaxy sample \cite{White:2011}  from  the SDSS-III Data Release 9 \cite{Ahn:2012}. The data set in question covers  three regions of the sky at $z\sim 0.5$ with  total volume of  $\sim 10$ (h$^{-1}$ Gpc)$^{3}$.  The bounds that the authors found on the bias parameter $b_r$ by using the SDSS-III Data Release 9 sample appears to be $b_r< 0.033$ measured at the 95$\%$ confidence level. The constraint on the relative velocity bias parameter yields the systematic error of $0.57\%$ on the  measurements of the BAO peak position at redshift $z = 0.57$, which is negligible compared to the current observational error of $1.7\%$. Note that the sample includes many coherent patches with different values of $\vbc$, therefore the analysis should give an effect averaged over the Maxwell-Boltzman distribution of the velocities. 

The results described above \cite{Yoo:2013} were based on the Fourier space picture. Z. Slepian and D. Eisenstein (2014) \cite{Slepian:2014} took a different approach (motivated by D. Eisenstein et al. (2007) \cite{Eisenstein:2007}, who use configuration space to show that the BAO produce a robust peak in the correlation function) and developed a configuration space picture. They provide a clear explanation for why the BAO peak is shifted by  ${\bf\vbc}$ as well as a configuration-space template for fitting the three-point correlation function  to isolate $b_r$. The authors have shown that the full three-point correlation function has robust radial signatures of the relative velocity effect that are unique and cannot come from any other bias term. 

Another possible correlation of the high-redshift  supersonic relative motion   and the low-redshift population of galaxies may be the effect of  ${\bf \vbc}$ on the age of the oldest stars in present day galaxies. 
 Detection of five stars with exceptionally low metal abundances (more precisely with metallicities\footnote{The metallicity $Z$ is often expressed  in terms of relative abundances of iron to hydrogen $Z  \equiv  \left(\frac{N_{Fe}}{N_{N_H}}\right)$. } below  $Z = 10^{-4.5}Z_\odot$) found within the  Milky Way \cite{Christlieb:2002, Frebel:2005, Norris:2007, Caffau:2011, Keller:2014} indicates that some of the stars in our galaxy   formed out of pristine gas possibly  enriched by at most a few low-energy supernovae. Note that metallicities lower than the critical value $Z_{crit} = 10^{-4}Z_{\odot}$ are attributed to  Population III stars, i.e., the first generation of stars in the history of the Universe. Possibly, the most remarkable discovery of this kind is that of a star with absolutely no traces of heavy elements in its optical spectrum  (the upper limit on metallicity of this star is  $Z = 10^{-7.1}Z_\odot$). The star,   found recently in the continuing Sky Mapper Southern Sky Survey \cite{Keller:2007}, most likely formed from primordial gas enriched by only one supernova with an original mass of about 60 M$_\odot$. Such extremely metal-poor stars are argued to be almost as old as the Universe, however their exact  ages were not measured by observations. Stars of this kind pose sometimes very hard puzzles to solve, as in the case of another low metallicity  $Z\sim 10^{-2.4} Z_\odot$ star found by the Hubble Space Telescope in our Galaxy the age was estimated to be $14.46\pm 0.31$ Gyr old \cite{Bond:2013}. The estimated age is actually in tension with the  age of the Universe determined from the latest CMB data \cite{Ade:2013}, namely $13.798\pm0.037$ Gyr (where the error bars indicate $68\%$ confidence level).

The formation of so old stars can naturally be affected by the presence of supersonic relative motion. However, whether or not a careful treatment of ${\bf \vbc}$ can resolve the tension between the age of the oldest observed stars and the age of the Universe is still an open question. What one can say with confidence is that   the relative motion may have an impact on massive low-redshift galaxies by affecting the population of the oldest stars in them (which presumably formed at very early stages of the cosmic history and thus could have been affected by ${\bf \vbc}$), and in particular, that the age of the oldest stars in such  galaxies should fluctuate on large scales  with the characteristic scale identical to that of BAO. Using the estimates  we discussed in section \ref{Sec:FS} for the delay in the formation of the very first star introduced by ${\bf \vbc}$, we can estimate the order of magnitude of the age  fluctuation for the oldest stars. For instance, the population of stars with metallicities lower than the critical value $Z_{crit}$ in galaxies located in regions with  $\vbc = 0$  can be up to $\sim 4$ Myr older than the most ancient stars in  galaxies within  patches with $\vbc = 1\sigma_{bc}$.

\subsection{Primordial non-Gaussianity versus supersonic motion}

Initial conditions for structure formation are often assumed to have Gaussian statistics,  in agreement with the observations of the early Universe at highest available redshifts $\sim 1100$ \cite{Ade:2013b}. However, the  origin and character of the initial conditions as well as the physical phenomena that generated the initial fluctuations in energy density remain unknown. More and more observational evidences hint that the initial conditions were seeded during cosmological inflation \cite{BICEP:2014}, but are still far from selecting the particular inflationary scenario out of numerous models proposed through the years. A measurement  of non-Gaussianity would be a powerful  tool in constraining possible inflationary scenarios, as different models predict different types of non-Gaussianity.  One of the convenient ways to parameterize the level of (local) non-Gaussianity is by writing the primordial gauge-invariant gravitational potential $\Phi$ in the form $\Phi = \Phi_L+f_{NL}^{local}\left[\Phi_L^2-<\Phi_L^2>\right]$, by putting in evidence $\Phi_L$  the part of the  potential following Gaussian statistics. The constraints on this kind of non-Gaussianity from observations of the CMB by Planck satellite give $f_{NL}^{local} = 2.7\pm5.8$ \cite{Ade:2013b}, consistent with a purely Gaussian field of $f_{NL} = 0$.  
  
Before the severe constraints on the  $f_{NL}$ parameter by Plank satellite came out, the effect of $f_{NL}$ on structure formation properties and the baryon history was simulated for the first time \cite{Maio:2011b} using a high-resolution cosmological N-body hydrodynamical simulation which also included details of the primordial chemistry.   In this simulation also  effect of ${\bf \vbc}$ was included and its impact  on structure formation  was compared to that of the primordial non-Gaussianity. Comparing the $f_{NL}=0$ and the $f_{NL}=100$  cases for a fixed value of $\vbc$, one  notices that the effects of non-Gaussianities are respectively negligible or comparable to those of $\vbc$ (for $f_{NL} = 100$ and $\vbc = 1\sigma_{bc}$). This means that  in the realistic case of $-3.1< f_{NL}<8.5$ \cite{Ade:2013b} the effect of local non-Gaussianity on structure formation is expected to be completely negligible compared to that of ${\bf \vbc}$.

\subsection{Primordial magnetic fields}

Weak cosmic magnetic fields are  found in various astrophysical systems, including stars \cite{Donati:2009},  low- and high-redshift galaxies \cite{Wielebinski:2005, Kronberg:1992, Kronberg:1994, Fletcher:2011, Beck:2012, Bernet:2008},  clusters \cite{Clarke:2001, Bonafede:2010, Feretti:2012},  superclusters \cite{Xu:2006} and  even  voids \cite{Neronov:2010, Dolag:2011, Tavecchio:2010, Tavecchio:2011, Vovk:2012, Taylor:2011, Dermer:2011}. Magnetic fields in galaxies and clusters have intensities of a few $\mu$G \cite{Beck:1996, Widrow:2002, Vallee:2004} and their  origin  is still uncertain. According to current understanding, they can be sourced by weak pre-existing magnetic fields of unknown origin which later are amplified by different types of dynamical processes in the structures. Alternatively, the initial magnetic fields could be generated during the first stages of gravitational collapse  (see a recent review by R. Durrer and A. Neronov (2013) \cite{Durrer:2013} for a more detailed discussion). Another possibility to generate the initial seeds of present day magnetic fields is through a process called {\it Biermann battery}: it was demonstrated by L. Biermann (1950) \cite{Biermann:1950} that if a plasma has a rotational vortex-like motion, electric currents (and thus magnetic fields) will be generated.  Evolution of the magnetic field  ${\bf B}$, generated through  Biermann battery, in time in an expanding universe \cite{Naoz:2013b} is given by the relation
\begin{equation}
\label{eq:magnetic}
\frac{\partial }{\partial t}\left(a^2 {\bf B}\right) = -\frac{c k \bar T_K}{e}\nabla\delta_e\times\nabla\delta_T,
\end{equation}
where $e$ is the  electron charge, and $\delta_e$ is the fluctuations in the free electron fraction, $x_e$. In other words, the  magnetic field can be  generated with this mechanism only if gradients in temperature and electron fraction are not parallel to each other. 

\begin{figure}[t]
{\center
\includegraphics[width=3in]{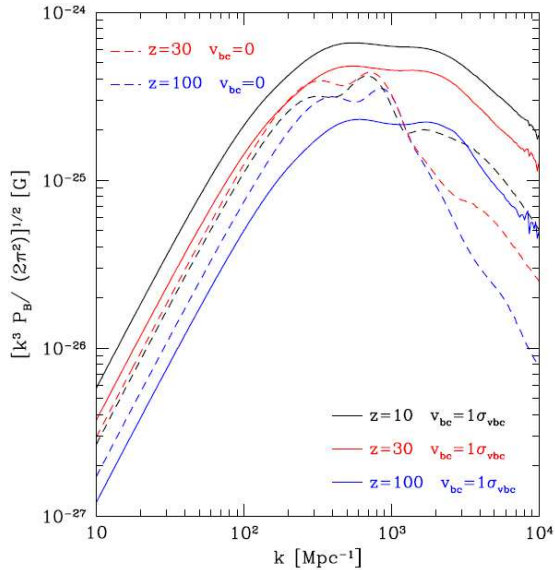}
\caption{\label{fig:mag}  Root mean square of the magnetic field generated by the Biermann battery as a function of wavenumber. Two cases are shown:  $\vbc = 0$ (solid lines), and a typical stream velocity, $\vbc = 1\sigma_{bc}$ (dashed lines). Three redshifts are considered: z = 100 (blue lines), z = 30 (red lines), z = 10 (black lines). The figure is adopted from S. Naoz and R. Narayan (2013) \cite{Naoz:2013b}.}}
\end{figure}

As was recently noted by  S. Naoz and R. Narayan (2013) \cite{Naoz:2013b}, the spatially varying speed of sound of baryons \cite{Naoz:2005} in the early Universe can produce the needed  vorticity in the residual free electrons, thus seeding a magnetic field of $10^{-19}-10^{-18}~\mu$G at redshifts of order a few tens.  This initial magnetic field is then expected to be  amplified after the end of cosmic reionization  (when gas heats up above the temperature of the CMB) to the values consistent with  observational constrains $\sim 10^{-12}~\mu$G \cite{Neronov:2010, Dermer:2011, Aleksic:2010}, and further increased by non-linear processes of structure formation at lower redshifts. The relative supersonic velocities can enhance the produced magnetic fields by adding an effective anisotropic pressure, and thus affecting the gradients in fluctuations of baryon temperature and electron fraction. Figure \ref{fig:mag},  adopted from  the paper by S. Naoz and R. Narayan (2013) \cite{Naoz:2013b}, shows the dependence of the root mean square magnetic field on scale, redshift and magnitude of the relative velocities. ${\bf \vbc}$ manifests itself most notably at small scales, where it enhances the rms of the magnetic field by a bit less than an order of magnitude at redshifts $z = 10-100$. Naturally, the enhancement is expected to be biased by the magnitude of local ${\bf \vbc}$, which imprints the BAO pattern in the primordial magnetic fields. If detected in the future, such modulation could be correlated with galaxy samples or 21-cm signal to single out the effect of ${\bf \vbc}$. 

\section{Final Remarks}

In this review we have overviewed the phenomenon of the supersonic relative velocity motion between baryons and dark matter. Since 2010, when the importance of the relative velocities was recognized for the first time \cite{Tseliakhovich:2010}, their impact on various areas of astrophysics and cosmology has been widely explored. The relative velocities were shown to leave  signatures in a surprisingly wide variety of cosmological probes, as they were shown to affect abundance \cite{Tseliakhovich:2010,  Maio:2011, OLeary:2012, Naoz:2012, Fialkov:2012, Bovy:2013, Tseliakhovich:2011,  Tanaka:2013,  Tanaka:2013b} and gas content \cite{  Maio:2011, Greif:2011,  OLeary:2012, Fialkov:2012, Naoz:2012,Naoz:2013, Tseliakhovich:2011, Dalal:2010, Richardson:2013, Naoz:2011} of $10^5-10^6$ M$_\odot$ halos at high redshifts (up to $z\sim10$), formation of the first stars \cite{ Maio:2011, Stacy:2011, Greif:2011, OLeary:2012,Fialkov:2012}, formation of globular clusters \cite{Naoz:2014}, abundance of  massive  black holes \cite{Tanaka:2013, Tanaka:2013b, Latif:2013, Visbal:2014}, generation of  primordial magnetic fields \cite{Naoz:2013b}, brightness of  satellite galaxies \cite{Bovy:2013}, clustering of  massive low-redshift galaxies \cite{Yoo:2011, Yoo:2013, Slepian:2014},  B-mode polarization of the CMB \cite{Ferraro:2012}, patchiness of reionization \cite{Bittner:2011, Su:2011}, and fluctuations in 21-cm signal of neutral hydrogen \cite{ Visbal:2012,Fialkov:2013, Dalal:2010, McQuinn:2012,  Fialkov:2014c}.
 
Being coherent on scales of few Mpc, the supersonic flows relatively are easy to include in small scale simulations of early structure formation. In addition to density fluctuations, they represent a supplementary set of initial conditions for the growth of cosmic structure, which should not be omitted especially when structure formation on small scales and high redshifts is explored. In the presence of  relative velocities, the standard linear theory where every wavenumber evolves separately is no longer valid \cite{Tseliakhovich:2010} due to the fact that  the evolution of  density fluctuations on small-scale is coupled to large-scale velocity modes. Omitting the effect of the supersonic relative motion  results in simulating an unrealistic patch of the universe, since the probability to find such a patch is negligibly low. In the case of large-scale cosmological calculations, properly including the velocities increases stochasticity in various astrophysical and cosmological quantities, adding a scale-dependent bias with a characteristic scale of order $150$ Mpc \cite{Fialkov:2012, Visbal:2012,  Fialkov:2013, Dalal:2010, McQuinn:2012}. In addition, for statistically averaged quantities, such as mean star formation rates,  the effect of relative velocities frequently does not average out. In total, the large scale relative velocities  should be included into the standard set of initial conditions for high-redshift cosmological simulations.  On the other hand, it should be kept in mind that processes such as the negative feedback by Lyman-Werner photons may lessen the significance of the velocity effects at redshifts  $z\lesssim 30$ \cite{Fialkov:2013, Visbal:2014}.  However, the exact interplay between feedbacks and velocities is yet uncertain.

Beside the effects introduced by modification of the linear theory,  the relative velocities  induce very interesting dynamics at small scales \cite{Maio:2011, Stacy:2011, Greif:2011,  OLeary:2012, Naoz:2014, Richardson:2013}. In addition to hindering star formation \cite{Maio:2011,Stacy:2011, Greif:2011,  OLeary:2012}, they perturb the accretion of gas into halos so that the process proceeds in an asymmetric manner; filaments of the accreting gas are disrupted more easily if they are perpendicular to the local direction of the relative velocity, thus increasing stochasticity of the gas fraction - halo mass relation. Moreover, halos that form in regions of relatively high velocity develop supersonic Mach cones as they move supersonically fast through the sea of baryons \cite{OLeary:2012, Richardson:2013}.

Various data sets, some of which already available today, are expected to bear the imprints of the relative velocities. Unaccounted for, the streaming motion is prone to bias parameters estimated from these data sets, as was discussed here in the context of the shift in BAO peak position found from the distribution of massive low-redshift galaxies \cite{Yoo:2011, Yoo:2013,  Slepian:2014} (section \ref{Sec:LRG}). It may be hard to extract the effects of the relative velocities from a single data set, especially when a weak dependence is expected. If this is the case,  correlating various data sets may dramatically improve the  chances to statistically detect and account for the velocity-induced signatures. Using a complementary data set may become particularly useful when the characteristic imprints of velocities in a specific cosmological probe are degenerate with traces of other astrophysical phenomena, such as radiative feedback. Techniques of this kind  will be essential if precise constrains on astrophysical parameters are to be achieved.

Finally, the stochasticity introduced by the relative velocities may propagate to cosmological signals which have not yet been  considered in the literature. For instance, it may affect predictions for the 21-cm forest, bias the distribution of high-redshift $\gamma$-ray sources, imprint BAOs in the pattern of metal enrichment (which in turn may affect the course of reionization), etc. What we have presented here makes clear that providing a  complete theoretical description  of the impact of ${\bf \vbc}$ on cosmic signals  is essential today, in the era of precision cosmology.

\section*{Acknowledgments}

Preprint of an article submitted for consideration in Int. J. Mod. Phys. D, $\copyright$ 2014 World Scientific Publishing Company, http://www.worldscientific.com/worldscinet/ijmpd.

The author would like to thank R. Barkana and S. Naoz for their essential comments regarding this manuscript. The author was supported   by the LabEx ENS-ICFP: ANR-10-LABX-0010/ANR-10-IDEX-0001-02 PSL and NSF grant AST-1312034 while working on this review.

{}
\end{document}